\documentclass[final,5p,times,twocolumn]{ms}

\usepackage{amssymb,amsfonts,amsmath,amstext,amsgen,amsopn,amsxtra,indentfirst,lscape}

\journal{the Publications of the Astronomical Society of Australia (PASA)}

\begin{document}

\begin{frontmatter}

\title{Balmer-Dominated Shocks: A Concise Review}

\author{Kevin Heng}
\ead{heng@ias.edu}

\address{Frank \& Peggy Taplin Member \\ Institute for Advanced Study, School of Natural Sciences, \\ Einstein Drive, Princeton, NJ 08540, U.S.A.}

\begin{abstract}
A concise and critical review of Balmer-dominated shocks (BDSs) is presented, summarizing the state of theory and observations, including models with/without shock precursors and their synergy with atomic physics.  Observations of BDSs in supernova remnants are reviewed on an object-by-object basis.  The relevance of BDSs towards understanding the acceleration of cosmic rays in shocks is emphasized.  Probable and possible detections of BDSs in astrophysical objects other than supernova remnants, including pulsar wind nebulae and high-redshift galaxies, are described.  The case for the continued future of studying BDSs in astrophysics is made, including their relevance towards understanding electron-ion temperature equilibration in collisionless shocks.
\end{abstract}


\end{frontmatter}

\section{Introduction}
\label{sect:intro}

\begin{figure}
\begin{center}
\vspace{-0.1in}
\includegraphics[width=\columnwidth]{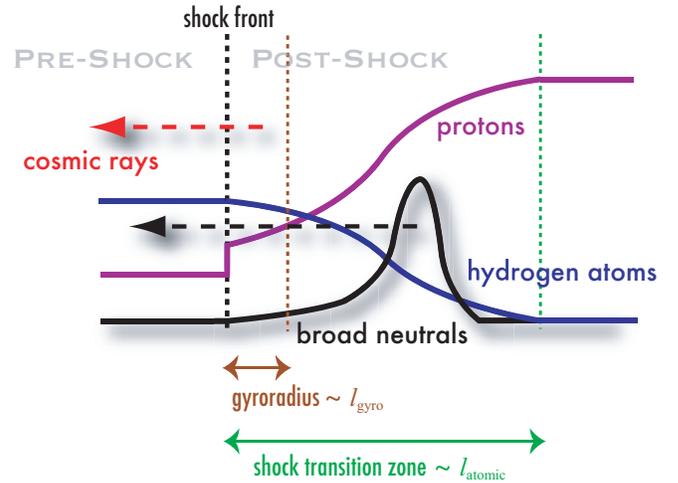}
\end{center}
\vspace{-0.3in}
\caption{Schematic structure of a Balmer-dominated shock.  As the cold, pre-shock hydrogen atoms stream past the collisionless shock front, they interact with the post-shock protons via charge transfer reactions to create a secondary population of atoms known as ``broad neutrals'' (\S\ref{subsect:atomic}).  The post-shock electrons and protons are heated in a thin layer with thickness $\sim l_{\rm gyro}$ (equation [\ref{eq:l_gyro}]).  Broad neutrals carry information about the post-shock gas and are created in the post-shock region known as the ``shock transition zone'' (\S\ref{subsect:models}) with a thickness $\sim l_{\rm atomic}$ (equation [\ref{eq:l_atomic}]).  The densities of the pre-shock hydrogen atoms, broad neutrals and protons are shown; the density jump for the protons reflects the Rankine-Hugoniot conditions.  The dashed arrows indicate that cosmic rays and broad neutrals may diffuse upstream and heat the pre-shock gas (\S\ref{subsect:models_precursors}).}
\vspace{0.1in}
\label{fig:schematic}
\end{figure}

Shocks are present wherever matter is accelerated past the sound speed of the medium they are propagating through \cite{zr66,mh80,dm93}.  Their ubiquity precludes any one author from performing an exhaustive review of their study, which includes theory, observation, experiment and simulation --- both terrestrial and astrophysical.  In this review, we focus on a specific class of astrophysical shocks known as ``Balmer-dominated shocks'' (BDSs; Figure \ref{fig:schematic}), previously reviewed by Raymond \cite{ray91,ray01}.  Traditionally observed around historical supernova remnants (SNRs), these are $\sim 200$--$9000$ km s$^{-1}$ shocks impinging upon the $\sim 0.1$--1 cm$^{-3}$ interstellar medium (ISM), manifesting themselves as faint (but limb-brightened), optical filaments.

\begin{figure}
\begin{center}
\includegraphics[width=0.95\columnwidth]{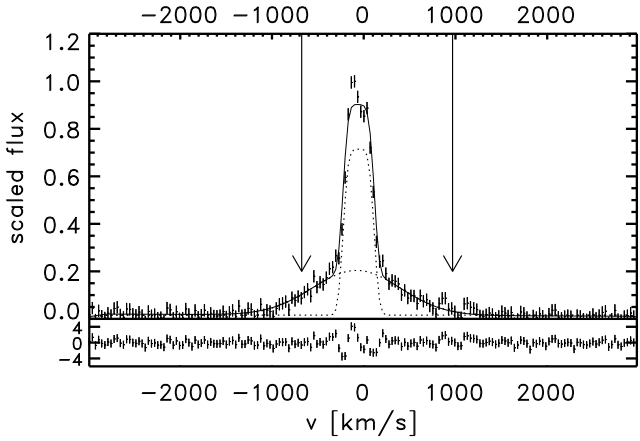}
\end{center}
\vspace{-0.2in}
\caption{Example of the two-component H$\alpha$ line from a Balmer-dominated shock located in northeastern RCW 86, a Galactic supernova remnant.  The vertical arrows indicate [N~{\sc ii}] $\lambda 6548, 6584$ emission.  Courtesy of Eveline Helder.}
\label{fig:halpha_line}
\end{figure}

BDSs are characterized by:
\begin{itemize}

\item Having relatively high velocities ($\gtrsim 200$ km s$^{-1}$) such that they are ``non-radiative'' (\S\ref{sect:physics});

\item Presence of strong hydrogen lines with narrow ($\sim 10$ km s$^{-1}$) and broad ($\sim 1000$ km s$^{-1}$) components (e.g., Figure \ref{fig:halpha_line});

\item Absence or weak presence of forbidden lines of lowly-ionized metals;

\item General lack of non-thermal X-ray emission at the locations where two-component H$\alpha$ lines are detected (but see \S\ref{sect:cosmic_rays}).

\end{itemize}
That the forbidden metal lines are weak relative to the hydrogen (and helium) lines is simply a consequence of the high temperatures in BDSs allowing the production of both $\gtrsim 10$ eV and $\lesssim 1$ eV lines --- the relative line strengths then scale as the elemental abundance --- in contrast to the conditions in emission line nebulae where the hydrogen lines are suppressed because of low temperatures.  BDSs are somewhat simple to interpret, because radiative cooling and recombination are unimportant in the immediate post-shock region.  

Combined with proper motion measurements, BDSs can be used to infer distances\footnote{It can be argued that, at least for SNRs, this method is an improvement over using the $\Sigma$-$D$ relation to obtain distances \cite{green91,smith97}.} to spatially-resolved objects.  If both the narrow and broad components are detected, the ratio of electron-to-ion temperatures just behind the shock front can be inferred \cite{ghava07a,v08}, thereby providing constraints for plasma physics simulations of electron-ion\footnote{In this paper, we use ``ions'' to mean non-leptonic ions, i.e., excluding electrons.} equilibration.  BDSs can be used to extract information on cosmic ray ions and study particle acceleration in partially neutral media.  The broad hydrogen lines produced generally serve as direct tracers of the post-shock proton distribution.  Understanding the physics of these fast, spatially-resolved shocks allows us to better understand if shocks are important in distant, spatially-unresolved, astrophysical objects.

In \S\ref{sect:physics}, we develop some intuition for the physical conditions in BDSs by familiarizing the reader with order-of-magnitude time and length scales.  In \S\ref{sect:history}, we gain historical perspective by tracking the pioneering efforts to study BDSs in SNRs.  In \S\ref{sect:snr}, we compile and summarize the SNR observations on an object-by-object basis.  The relevance of BDSs towards studying cosmic ray acceleration is emphasized in \S\ref{sect:cosmic_rays}.  In \S\ref{sect:theory}, we review the theoretical models of BDSs, including those with and without shock precursors; we also describe the atomic cross sections that are imperative for constructing these models.  Probable and possible observations of BDSs in other astrophysical objects are described in \S\ref{sect:others}.  The case for the continued future of studying BDSs is made in \S\ref{sect:future}.  Table \ref{tab:symbols} contains a list of symbols that are commonly used throughout the paper.

\begin{table*}
\centering
\begin{minipage}{140mm}
\label{tab:symbols}
\begin{tabular}{ll}
\hline\hline
\multicolumn{1}{c}{Symbol} & \multicolumn{1}{c}{Meaning}\\
\hline
\vspace{2pt}
$v_s$ & shock velocity\\
$\Delta v$ & relative velocity between atoms and ions\\
$n$ & ambient density (total)\\
$f_{\rm ion}$ & pre-shock ion fraction\\
$B_{\rm ISM}$ & magnetic field strength of ambient ISM\\
$T_0$ & pre-shock temperature\\
$T_s$ & post-shock temperature\\
$T_e, T_p, T_i$ & post-shock electron, proton and ion temperatures\\
$\beta$ & ratio of electron to proton temperatures\\
$W_0$ & full width at half-maximum of narrow hydrogen$^\dagger$ line\\
$W$ & full width at half-maximum of broad hydrogen$^\dagger$ line\\
$I_b/I_n$ & ratio of broad to narrow hydrogen$^\dagger$ line intensities\\
\hline
$\dagger$: typically H$\alpha$ unless otherwise noted.
\end{tabular}
\end{minipage}
\caption{Commonly-used symbols}
\end{table*}

\section{Order-of-Magnitude Physics: Time and Length Scales}
\label{sect:physics}

A shock with a velocity $v_s$ impacting upon ambient ISM with a density $n$, magnetic field strength $B_{\rm ISM}$ and average gas mass $\mu$ has sonic and Alfv\'{e}nic\footnote{We differentiate between Alfv\'{e}nic and magnetosonic Mach numbers; the latter is defined as $v_s$ relative to the sound and Alfv\'{e}n speeds summed in quadrature.} Mach numbers of
\begin{equation}
{\cal M}_s \sim v_s \sqrt{\frac{\mu}{kT_0}} \sim 100  ~\left(\frac{v_s}{1000 \mbox{ km s}^{-1}} \right) \left(\frac{\mu}{m_{\rm H}} \right)^{1/2} \left(\frac{T_0}{10^4 \mbox{ K}}\right)^{-1/2} 
\end{equation}
and
\begin{equation}
\begin{split}
{\cal M}_A &\sim \frac{2v_s \sqrt{\pi n \mu}}{B_{\rm ISM}} \\
&\sim 500 ~\left(\frac{v_s}{1000 \mbox{ km s}^{-1}} \right) \left(\frac{n}{1 \mbox{ cm}^{-3}} \frac{\mu}{m_{\rm H}} \right)^{1/2} \left(\frac{B_{\rm ISM}}{1 ~\mu\mbox{G}} \right)^{-1},
\end{split}
\label{eq:alfven_mach}
\end{equation}
implying that high-velocity shocks are inaccessible to terrestrial investigations, where $m_{\rm H}$ is the mass of the hydrogen atom and $k$ is the Boltzmann constant.

If the pre-shock hydrogen gas has a temperature $T_0$ and emits a thermal, Gaussian-shaped line, its full width at half-maximum (FWHM) is
\begin{equation}
W_0 = 2 \sqrt{\frac{2 \ln{2} ~k T_0}{m_{\rm H}}} = 21 \mbox{ km s}^{-1} ~\left(\frac{T_0}{10^4 \mbox{ K}}\right)^{1/2}.
\label{eq:narrow_width}
\end{equation}
Since $T_0 \approx 10^4$ is the temperature above which collisional ionization overwhelms recombination (e.g., see Figure 12.4 of Osterbrock \& Ferland \cite{osterbrock06}), narrow hydrogen lines with widths greater than about 20 km s$^{-1}$ are believed to be non-thermally broadened (see \S\ref{subsect:broad_narrow_line}).  In this paper, we take the thermal velocity of an ion to be the root mean square (rms) velocity in three dimensions, given by
\begin{equation}
v_{{\rm th},i} = \sqrt{\frac{3kT_i}{m_i}},
\label{eq:vthermal}
\end{equation}
where $T_i$ and $m_i$ are the ion temperature and mass, respectively.  The thermal velocity of an electron is obtained by inserting $T_e$ and $m_e$ into equation (\ref{eq:vthermal}).

The shock heats up the impacted material to a temperature of
\begin{equation}
T_s = \frac{2\left(\gamma-1\right)}{\left(\gamma+1\right)^2} \left(\frac{m_i}{k} \right) v^2_s = 2.3 \times 10^7 \mbox{ K} ~\left(\frac{m_i}{m_p}\right) \left(\frac{v_s}{1000 \mbox{ km s}^{-1}} \right)^2,
\label{eq:tshock}
\end{equation}
where $\gamma$ is the adiabatic index of the shocked gas and $m_p$ is the proton mass.  The second equality assumes $\gamma=5/3$, i.e., a monoatomic, non-relativistic gas.
Emission lines from the post-shock gas typically have a width of
\begin{equation}
W \sim \frac{4v_s}{\gamma+1} \sqrt{\ln{2} ~\left(\gamma - 1\right)}.
\label{eq:broad_width}
\end{equation}
For hydrogen lines (see \S\ref{sect:history}), there is a non-negligible dependence of $W$ on $\beta \equiv T_e/T_p$, the ratio of electron-to-proton temperatures \cite{hm07,v08}.  For $\gamma = 5/3$, we have $W \sim v_s$.

For lack of a better term, BDSs are described as being ``non-radiative'', meaning that the age of the shock is less than both the cooling and recombination time scales in the intermediate post-shock region,
\begin{equation}
t_{\rm age} < \mbox{min}\left\{t_{\rm cool},t_{\rm rec}\right\}.
\label{eq:nonradiative}
\end{equation}
In other words, the shock does not produce radiation that is dynamically important.

Accurate determinations of the cooling time scale requires specifying the elemental abundance of --- and calculating the radiative transfer of photons through --- the post-shock plasma \cite{sd93}, but a simple estimate can be obtained using
\begin{equation}
t_{\rm cool} \sim \frac{kT_s}{n \Lambda} \approx 4 \times 10^5 \mbox{ yrs} ~\left(\frac{T_s}{10^7 \mbox{ K}}\right) \left(\frac{n}{1 \mbox{ cm}^{-3}} \frac{\Lambda}{10^{-22} \mbox{ erg} \mbox{ cm}^3 \mbox{ s}^{-1}} \right)^{-1},
\end{equation}
where $T_s$ is the post-shock temperature, $n$ is the ambient density and $\Lambda$ is the cooling function.

Recombination becomes important when the post-shock plasma cools to $\sim 10^4$ K; the recombination time scale is approximately
\begin{equation}
t_{\rm rec} \sim \frac{1}{n \alpha_{\rm rec}} \approx 8 \times 10^4 \mbox{ yrs} ~\left(\frac{n}{1 \mbox{ cm}^{-3}}\right)^{-1} \left(\frac{T_s}{10^4 \mbox{ K}}\right)^{0.7},
\label{eq:recombination}
\end{equation}
where $\alpha_{\rm rec}$ is the recombination coefficient \cite{hui97, osterbrock06}.  More accurate expressions for $\alpha_{\rm rec}$ are given in \cite{vf96}.

It is apparent that for shocks with ages $\sim 100$--1000 years and velocities $\sim 1000$ km s$^{-1}$ impinging upon the ISM with $\sim 0.1$--1 cm$^{-3}$ densities, the non-radiative condition described in equation (\ref{eq:nonradiative}) is trivially fulfilled.

BDSs are also collisionless shocks, because the length scale for atomic interactions  at $\sim 1000$ km s$^{-1}$ (typically dominated by charge transfer reactions at these velocities; see \S\ref{subsect:atomic}),
\begin{equation}
l_{\rm atomic} \sim \frac{1}{n \sigma_{\rm atomic}} \sim 10^{15} \mbox{ cm} ~\left(\frac{n}{1 \mbox{ cm}^{-3}}\right)^{-1},
\label{eq:l_atomic}
\end{equation}
is much larger than reasonable estimates for the proton gyroradius,
\begin{equation}
l_{\rm gyro} \sim 10^{10} \mbox{ cm} ~\left(\frac{v_p}{1000 \mbox{ km s}^{-1}} \right) \left(\frac{B_{\rm ISM}}{1 ~\mu\mbox{G}} \right)^{-1},
\label{eq:l_gyro}
\end{equation}
where $v_p$ is the velocity of the proton.  The corresponding gyroradius for an electron is smaller by a factor of $m_e/m_p$, where $m_e$ is the electron mass.

The condition $l_{\rm gyro} \ll l_{\rm atomic}$ implies that the relative temperatures of the electrons and ions are determined by collective electromagnetic interactions, rather than by physical collisions. The time scale for Coulomb equilibration \cite{spitzer62,fl97} is
\begin{equation}
t_{\rm ei} \sim 3 \times 10^{4} \mbox{ yrs} ~\left(\frac{T_e}{10^7 \mbox{ K}}\right) \left(\frac{n}{1 \mbox{ cm}^{-3}} \frac{{\cal C}}{10}\right)^{-1},
\label{eq:tei}
\end{equation}
where $T_e$ is the electron temperature and ${\cal C}$ is the Coulomb logarithm.  Since $t_{\rm ei} \gg t_{\rm age}$, Coulomb interactions alone cannot bring about temperature equilibration in young SNRs (with ages $\sim 10^3$ yrs).  The range of electron-to-ion temperature ratios is expected to be $m_e/m_i \le T_e/T_i \le 1$ \cite{zr66}.

In the absence of limb brightening, the intensity or brightness of a shock filament can be used to provide an estimate for the ambient density, since
\begin{equation}
\begin{split}
I_{\rm H\alpha} &\sim \frac{n v_s \epsilon_{\rm H\alpha} E_{\rm H\alpha}}{4\pi} \\
&= 5 \times 10^{-6} \mbox{ erg cm}^{-2} \mbox{ s}^{-1} \mbox{ sr}^{-1} ~\left(\frac{n}{1 \mbox{ cm}^{-3}} \frac{v_s}{1000 \mbox{ km s}^{-1}} \frac{\epsilon_{\rm H\alpha}}{0.2}\right),
\end{split}
\end{equation}
where $\epsilon_{\rm H\alpha}$ is the number of H$\alpha$ photons per ionization event and $E_{\rm H\alpha} \approx 2$ eV is the energy of a H$\alpha$ photon.

Energetic particles and/or photons may be generated in the immediate post-shock region of a non-radiative shock.  When the typical propagation velocities of the energetic particles exceed the shock velocity --- a condition automatically satisfied by the photons --- they will travel upstream into the pre-shock gas and heat it, thus broadening\footnote{The pre-shock hydrogen atoms affected by the precursor must still cross the collisionless shock front to produce the observed narrow lines.} the line width $W_0$.  Such broadening mechanisms are known as ``precursors'' and may be responsible for the unusually large values of $W_0$ observed in several SNRs (see \S\ref{subsect:broad_narrow_line}).  In shocks where a cosmic ray precursor is believed to be present, the length scale for cosmic ray propagation ($l_{\rm CR} \sim \kappa_{\rm CR}/v_s$) must be shorter than that for ionization, thereby setting an upper limit to the cosmic ray diffusion coefficient:
\begin{equation}
\kappa_{\rm CR} \lesssim 10^{24} \mbox{ cm}^2 \mbox{ s}^{-1} ~\left(\frac{n}{1 \mbox{ cm}^{-3}}\right)^{-1}  \left(\frac{v_s}{1000 \mbox{ km s}^{-1}} \right),
\label{eq:cr_coeff}
\end{equation}
where the typical cross section for ionization is taken to be $\sim 10^{-16}$ cm$^2$.  

\section{Historical Developments: the Remnants of Tycho, SN 1006 and the Cygnus Loop}
\label{sect:history}

In 1959, Minkowski \cite{minkowski59} announced, at the Paris symposium of the {\it International Astronomical Union}, the detection of bright H$\alpha$ filaments around the remnants of Kepler, Tycho and the Cygnus Loop\footnote{Minkowski also noted the presence of more than 200 ``small condensations'' in the nebulosity around Cassiopeia A.}.  Following an unsuccessful search for the optical remnant of supernova (SN) 1006 by Walter Baade, van den Bergh \cite{vdb76} in 1976 reported what he termed ``delicate wisps of filamentary nebulosity'' about $10^\prime$ from the center of the associated radio source.  He noticed that the morphology of the filaments were similar to those detected in Tycho's SNR (the remnant of SN 1572) and S147.  Shortly afterwards, Kirshner \& Taylor \cite{kt76} presented H$\alpha$ velocity measurements of up to 300 km s$^{-1}$ in the Cygnus Loop equivalent to shock temperatures $\sim 10^6$ K, consistent with the observation of soft X-rays and [Fe {\sc xiv}] $\lambda 5303$ emission behind the shock front.  

In 1978, McKee, Cowie \& Ostriker \cite{mco78} suggested that the high-velocity H$\alpha$ emission may be from pre-existing clouds in the SNR being accelerated and shock-heated by the blast waves, although they leave open the question of whether the clouds can survive in the environment of the SNR.  One prediction of the ``accelerated clouds'' model is that the H$\alpha$ emission should be accompanied by substantial emission in, for example, the [O {\sc i}] $\lambda 6300$ and [O {\sc iii}] $\lambda 5007$ lines.

In the same year, an optical spectrum of Tycho's SNR obtained by Kirshner \& Chevalier \cite{kc78} provided important clues towards understanding the emission mechanisms associated with the filaments.  Modeling the thermal X-ray continuum measured by {\it Ariel 5} yielded emission components with two different temperatures, 0.55 and 3.5 keV.  Assuming a distance to Tycho's SNR of 6 kpc, the tangential velocity of the filaments was estimated to be 5600 km s$^{-1}$, corresponding to post-shock temperatures near 40 keV.  A natural way to reconcile these two observables was for incomplete temperature equilibrium between the electrons and ions to occur --- in this case, for $T_i/T_e \sim 20$--30 such that the electron temperature is $T_e \sim 5 \times 10^5$ K.

A second important clue provided by the observations of Kirshner \& Chevalier was that only H$\alpha$ and H$\beta$ emission lines were observed over the wavelength range of 4700--8000 \AA.  Forbidden lines of singly- or doubly-ionized metals (e.g., [O {\sc iii}] $\lambda 4959, 5007$) were conspicuously absent, thereby precluding their use as density or temperature diagnostics and placing into question the accelerated clouds model.  Two interpretations were suggested: either the forbidden lines were collisionally quenched due to high densities, or the high temperatures caused all of the metals to be in their highest ionization states.  The absence of the [O {\sc iii}] $\lambda 5007$ line and the [O {\sc ii}] $\lambda 7319, 7329$ doublet requires densities in excess of $10^7$ cm$^{-3}$, orders of magnitude above typical ISM densities, thus ruling out the former possibility.

Also in 1978, Schweizer \& Lasker \cite{sl78} obtained spectra in the wavelength range of 3650--7500 \AA~ for BDSs at the northwestern rim of the remnant of SN 1006.  Only Balmer emission lines were detected; the [O {\sc i}], [O {\sc ii}], [O {\sc iii}], [N {\sc ii}] and [S {\sc ii}] lines were either absent or weak ($\lesssim 8\%$ by intensity relative to H$\alpha$ when uncorrected for reddening).

An alternative model was proposed by Chevalier \& Raymond \cite{cr78}, in which a collisionless shock impacts partially neutral gas.  Their model was able to explain the intensity, spectrum and width of the filaments observed in Tycho's SNR, including the weakness of the forbidden metal lines.  Lines of neutral helium were predicted to be the next brightest optical lines after the hydrogen Balmer lines; for example, the He~{\sc i} $\lambda 5876$ line is expected to be about 10 times fainter than the H$\beta$ line.  Limb brightening increased the intensity at the edge of the SNR by a factor $\sim (R_{\rm SNR}/l_{\rm ion})^{1/2}$, where $R_{\rm SNR}$ is the radius of the remnant and $l_{\rm ion} \sim 10^{16}(1\mbox{ cm}^{-3}/n)$ cm is the ionization length scale.  Bychkov \& Lebedev \cite{bl79} estimated the effect of photoionization of the pre-shock gas by photons from the post-shock gas; they predicted that high-velocity hydrogen emission should be expected from Kepler's SNR, but not from nebulae surrounded by H~{\sc ii} regions (e.g., Cas A, Crab).

\begin{figure}
\begin{center}
\includegraphics[width=0.9\columnwidth]{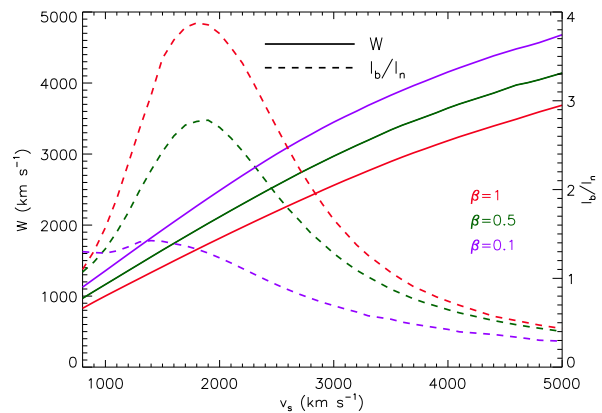}
\end{center}
\vspace{-0.3in}
\caption{Diagnostics for edge-on shocks using the H$\alpha$ line, based on calculations from van Adelsberg et al. \cite{v08}.  The width of the broad line $W$ and the ratio of the broad to narrow line intensities $I_b/I_n$ both depend on the shock velocity $v_s$ and the ratio of electron to proton temperatures $\beta$.  For illustration, we adopt Case A and B conditions for the broad and narrow lines (see \S\ref{subsect:atomic}), respectively, and fix the pre-shock ion fraction at $f_{\rm ion} = 0.5$.}
\label{fig:diagnostics}
\end{figure}

In a seminal 1980 paper, Chevalier, Kirshner \& Raymond \cite{ckr80} reported the detection of H$\alpha$, H$\beta$ and H$\gamma$ lines in Tycho's SNR and also expanded upon the shock wave theory proposed by Chevalier \& Raymond.  The H$\alpha$ line appeared to have two components: an unresolved, narrow component due to direct excitation of cold, pre-shock hydrogen atoms; and a broad component with a FWHM of $W=1800 \pm 200$ km s$^{-1}$.  The broad line component arises from a secondary population of hydrogen atoms created from charge transfer reactions between cold, pre-shock atoms and hot, post-shock protons that are called ``broad neutrals''.  \emph{Thus, broad neutrals carry valuable information about the post-shock gas.}

The Chevalier, Kirshner \& Raymond paper established several important results.  Firstly, $W$ is a diagnostic for the shock velocity (equation [\ref{eq:broad_width}]).  Secondly, the ratio of broad-to-narrow line intensities, denoted by $I_b/I_n$, is a sensitive function of the shock velocity, because the charge transfer cross section depends strongly on the electron and proton velocities (see \S\ref{subsect:atomic}).  Both $W$ and $I_b/I_n$ depend on the degree of temperature equilibration between electrons and ions (see Figure \ref{fig:diagnostics}).  Thirdly, it was realized that the conversion of Lyman to Balmer photons via scattering, known as ``Lyman line trapping'' (see \S\ref{subsect:atomic} for details), is a non-negligible effect in Tycho's SNR.  In pre-shock media where the optical depth to Ly$\beta$ photons equals or exceeds unity, one H$\alpha$ photon is produced for every 5--10 atoms that are ionized, a result that is confirmed by later calculations \cite{michael03,hm07}.  Combining the inferred value of the shock velocity with proper motion measurements, the distance to Tycho's SNR was revised downward to $2.3 \pm 0.5$ kpc.

\section{Observations of Balmer-Dominated Shocks in Historical Supernova Remnants}
\label{sect:snr}

\begin{figure}
\begin{center}
\includegraphics[width=0.8\columnwidth]{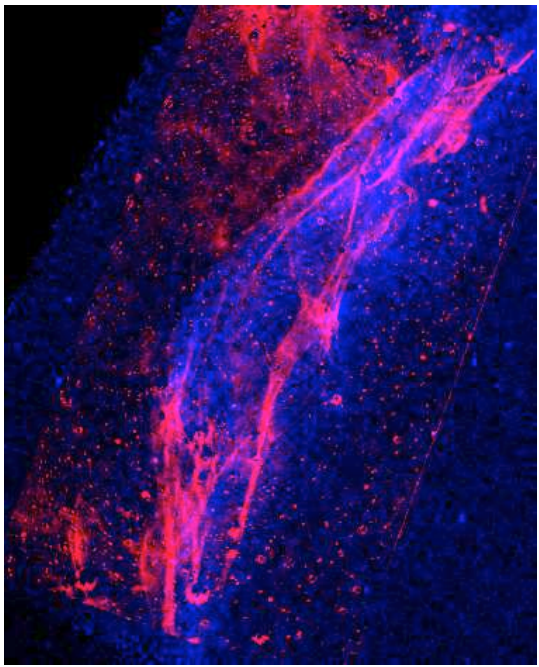}
\end{center}
\vspace{-0.2in}
\caption{Combined optical and X-ray image of northeastern RCW 86 (a Galactic supernova remnant), adapted from Helder et al. \cite{helder09}.  The blue and pink colors represent the {\it Chandra} broad-band (0.1--10 keV) and {\it VLT} narrow-band H$\alpha$ emission, respectively.  Courtesy of Eveline Helder.  (North is up and east is left.)}
\label{fig:rcw86}
\end{figure}

Following the pioneering studies described in \S\ref{sect:history}, the field of Balmer-dominated shocks grew out of optical observations made of faint shock filaments around historical SNRs\footnote{We refrain from using the term ``Balmer-dominated supernova remnants'', because a multitude of both non-radiative and radiative shocks may exist around a given remnant.} that predominantly have Type Ia (i.e., thermonuclear) SN progenitors.  A sample image of the shock filaments is shown in Figure \ref{fig:rcw86} for RCW 86.  Reported values of $W$ and $I_b/I_n$ are typically obtained from fitting Gaussians to the observed lines.

The translation of the measured $W$ and $I_b/I_n$ values (usually from H$\alpha$ lines) into reported values of $v_s$ is traditionally hampered by uncertainties in the temperature equilibration between electrons and ions.  Generally, the geometry of the ambient magnetic field in these SNRs is poorly known, particularly in remnants where radio emission is weak.

\subsection{Galactic SNRs}

\subsubsection{The Cygnus Loop}
\label{subsect:cygnus}

Raymond et al. \cite{ray83} report optical and ultraviolet observations of shock filaments at the eastern limb of the Cygnus Loop.  The two-component H$\alpha$ line is detected; by fitting Gaussians to each component, the narrow and broad FWHMs are measured to be 31 and 167 km s$^{-1}$, respectively.  The centers of the lines differ by less than the uncertainty in centering the broad component, indicating that the shock is moving almost exactly transverse to the line of sight, i.e., an edge-on shock.  A search for the variation of the broad H$\alpha$ line width across the shock filament proves to be unsuccessful.

Raymond et al. construct two models: the first assumes that plasma turbulence equilibrates the electron and ion temperatures (i.e., $T_e = T_i$) rapidly (compared to any time scale for atomic interactions); the second assumes only (slow) Coulomb equilibration, i.e., $T_e \ll T_i$.  They demonstrate (see their Table 2) that the models with only Coulomb equilibration match a series of line intensities, as well as the strength of the hydrogen two-photon continuum, better than the rapid equlibration models.  Their models also predict that bright O~{\sc vi} emission should be present in $\sim 200$ km s$^{-1}$ shocks.

Fesen \& Itoh \cite{fi85} measure the line intensities of H, He~{\sc i}, He~{\sc ii} and several low-ionization metals along a different portion of the same filament studied by Raymond et al.  By matching the theoretical and observed line intensities, they conclude that two- and three-temperature models fit the data better than the one-temperature model.  The three-temperature models are motivated by wishing to distinguish between shocked electrons and secondary electrons produced via ionization of neutral atoms.  Nevertheless, the three-temperature model does not provide significant improvements over the two-temperature one.  Curiously, the observed [N~{\sc ii}] $\lambda6583$ and [S~{\sc ii}] $\lambda6717, 6731$ line intensities exceed the model predictions by at least an order of magnitude, which motivate two interpretations: either the observed shock has a radiative component, or dust grains present in the non-radiative shock are reflecting emission from a proximate, radiative filament.

Ghavamian et al. \cite{ghava01} measure $W = 262 \pm 32$ km s$^{-1}$ and $I_b/I_n = 0.59 \pm 0.3$ in northeastern Cygnus.  Using their models (see \S\ref{subsect:models}), they find nearly complete ($\beta \approx 0.7$--1) electron-proton temperature equilibration based on the H$\beta$ line ($I_b/I_n = 0.99 \pm 0.3$), but are not able to match the measured $I_b/I_n$ value for H$\alpha$.

Salvesen, Raymond \& Edgar \cite{salvesen08} use proper motion measurements and X-ray spectra to constrain the ratio of cosmic ray to gas pressure behind BDSs in the northeastern rim.  Adopting a distance of $576 \pm 61$ pc to the Cygnus Loop, they derive shock velocities from the proper motion data.  Two-temperature fits to {\it ROSAT} spectra yield post-shock temperatures; the lower of the two temperatures is used to obtain an upper limit to the cosmic ray pressure.  Full equilibration between electrons and ions is assumed.\footnote{This is a conservative assumption for the purpose of obtaining upper limits to the cosmic ray pressure, which is dominated by the ions since they constitute the bulk of the fluid mass.}  For 18 non-radiative H$\alpha$ filaments, they find that the ratio of cosmic ray to gas pressure is close to zero.

\subsubsection{The Remnant of SN 1006}
\label{subsect:sn1006}

More than six years following an unsuccessful search by B.M. Lasker, Kirshner, Winkler \& Chevalier \cite{kwc87} report a convincing detection of the two-component H$\alpha$ line from a filament along the northwestern rim of SN 1006.  They measure $W = 2600 \pm 100$ km s$^{-1}$ and $I_b/I_n = 0.77 \pm 0.08$.  Bracketing the inferred shock velocities by models with no and full temperature equilibration between electrons and ions, and combining them with proper motion measurements, the distance to SN 1006 is estimated to be in the range 1.4--2.1 kpc.  Smith et al. \cite{smith91} later reanalyze the spectrum from Kirshner, Winkler \& Chevalier using improved line fitting techniques, thereby refining the broad H$\alpha$ line width to $2310 \pm 210$ km s$^{-1}$ and $I_b/I_n$ to $0.73 \pm 0.06$.

Laming et al. \cite{laming96} report estimates of the ratio of electron to proton temperatures behind a SNR shock front, by analyzing ultraviolet lines of He~{\sc ii} $\lambda 1640$, C~{\sc iv} $\lambda 1550$, N~{\sc v} $\lambda 1240$ and O~{\sc vi} $\lambda 1036$, and concluding that complete electron-ion equilibration is disfavored (see also \S\ref{subsect:equilibration}).

Ghavamian et al. \cite{ghava02} report the first detection of the He~{\sc i} $\lambda 6678$ line in a BDS, in addition to measuring the broad H$\alpha$ line width ($W=2290 \pm 80$ km s$^{-1}$) and a possible ($\sim 1.5\sigma$) detection of the He~{\sc ii} $\lambda 4686$ line.  The broad to narrow line ratio is measured for both H$\alpha$ ($I_b/I_n = 0.84^{+0.03}_{-0.01}$) and H$\beta$ ($I_b/I_n = 0.93^{+0.18}_{-0.16}$).  Based on their models (see \S\ref{subsect:models}), they infer $\beta \le 0.07$.

Raymond et al. \cite{ray07} use a deep {\it Hubble Space Telescope} image to resolve the H$\alpha$ filament in the northwest.  At 40 positions along the filament, they extract spatial profiles of the H$\alpha$ intensity across the shock front.  The exact details of the profile depend upon a combination of the pre-shock neutral density and neutral fraction --- thus, matching the profiles to models \cite{ray07,heng07} allows these pair of values to be determined at each position.  Taking the neutral fraction to be 0.05--0.2, Raymond et al. infer the pre-shock neutral density to be about 0.15--0.4 cm$^{-3}$ (see also \S5 of \cite{heng07}).  Also detected are parsec-scale ripples in the filament, which may reflect $\sim 20$\% density fluctuations caused by interstellar turbulence.

The distance to SN 1006 is a subject of controversy with quoted values ranging from 0.7 to 2.5 kpc; these estimates are anchored by the fact that spectra of the Schweizer-Middleditch star \cite{sm80}, located at $\sim 1.05$--2.1 kpc, exhibit absorption by SN 1006 ejecta \cite{burleigh00}.  Hamilton, Fesen \& Blair \cite{hamilton07} measure the expansion velocity of ejecta at the reverse shock to be $7026 \pm 10$ km s$^{-1}$, implying a shock radius of about 7 pc.  Since the remnant is approximately 15$^\prime$ across, this corresponds to a minimum distance of about 1.6 kpc.  The most direct determination\footnote{The caveat is that the main systematic uncertainty ($\sim 10$--30\%) is contained within the atomic cross sections (see \S\ref{subsect:atomic}).} is to combine measurements of the broad H$\alpha$ line with filamentary proper motions \cite{wl97}.  Van Adelsberg et al. \cite{v08} use their estimates of the shock velocity to revise the inferred distance from $\sim 2.2$ kpc to $\sim 1.6$ kpc, a somewhat more comfortable figure in comparison to that of the Schweizer-Middleditch star.

Overall, the remnant of SN 1006 has the distinction of being the only instance of a spatially-resolved SNR where both efficient and inefficient sites for particle acceleration have been identified in BDSs (see \S\ref{subsect:pneutral}), but yet shows a lack of evidence for precursor heating (\S\ref{subsect:broad_narrow_line}).

\subsubsection{Tycho's Remnant (SNR 1572)}
\label{subsect:tycho}

Kirshner, Winkler \& Chevalier \cite{kwc87} detect the two-component H$\alpha$ line from Tycho's SNR, measuring $W = 1800 \pm 100$ km s$^{-1}$ and $I_b/I_n = 1.08 \pm 0.16$, thereby providing improvements to the measurements of Chevalier, Kirshner \& Raymond and an estimate for the distance to the remnant of 2.0--2.8 kpc.

Smith et al. \cite{smith91} report significant variations in the H$\alpha$ line profile along the filament containing ``knot g'', an over-dense region located at the northeastern rim of the remnant.  The broad line width measured ($W = 1900 \pm 300$ km s$^{-1}$) is similar to that reported by Kirshner, Winkler \& Chevalier, but the broad-to-narrow line intensity ratio is somewhat lower ($I_b/I_n = 0.77 \pm 0.09$).  The lower limit of the distance to Tycho's remnant is revised to 1.5 kpc, with the full range (1.5--3.1 kpc) again being bracketed by uncertainties in the electron-ion temperature equilibration.

Ghavamian et al. \cite{ghava01} report $W = 1765 \pm 110, 2105 \pm 130$ km s$^{-1}$ and $I_b/I_n = 0.67 \pm 0.1, 0.75 \pm 0.1$ from two locations in knot g.  They are able to measure Balmer decrements (i.e., the H$\alpha$/H$\beta$ intensity ratio; see \S\ref{subsect:atomic0}) for both the narrow and broad lines, finding H$\alpha$/H$\beta \approx 4$--6 (narrow) and $\approx 3$--4 (broad).  The larger values of the Balmer decrements for the narrow lines indicate that Lyman line trapping, i.e., the conversion of Ly$\beta$ to H$\alpha$ and Ly$\gamma$ to H$\beta$, is relatively more important for the narrow compared to the broad lines.

Lee et al. \cite{lee07} present evidence for a photoionization precursor, with a width $\sim 10^{16}$ cm, in knot g --- He~{\sc ii} $\lambda 304$ photons in the post-shock region diffuse upstream and heat the pre-shock gas.  The narrow H$\alpha$ line has an unusually broad width ($\sim 45$ km s$^{-1}$), but the photoionization precursor is not believed to be the main cause of the broadening (see \S\ref{subsect:broad_narrow_line}).

\subsubsection{Kepler's Remnant (SNR 1604)}

Fesen et al. \cite{fesen89} discover BDSs along the northern limb of Kepler's remnant.  Averaging over BDSs observed at two slit locations, they find $W = 1750 \pm 200$ km s$^{-1}$ and $I_b/I_n = 1.1 \pm 0.25$.  The diffuse nature of the northern filaments is in contrast to the sharply-defined filaments observed in the remnants of Tycho and SN 1006.

Blair, Long \& Vancura \cite{blair91} scrutinize the remnant in greater detail and confirm the result of Fesen et al. --- diffuse BDS emission is present in the north indicative of the SN shock encountering lower density, uniform, pre-supernova material.  They also report the discovery of the first BDSs with knotty structures.  In general, the knots are categorized to be radiative, non-radiative and transitional; the transitional knots have weaker [S~{\sc ii}] emission compared to the bright, radiative knots, while the best examples of the non-radiative knots have no detectable [S~{\sc ii}] emission.  Adopting shock velocities of 1530--2000 km s$^{-1}$, the distance to Kepler's remnant is estimated to be $2.9 \pm 0.4$ kpc.

\subsubsection{RCW 86}

Generally --- but not universally --- accepted as the remnant of SN 185, the shell-like X-ray and radio source is found by Long \& Blair \cite{lb90} to contain BDSs ($W = 660 \pm 90$ km s$^{-1}$, $I_b/I_n = 1.0 \pm 0.2$).  Figure 3 of their study is particularly illuminating in that it shows sample spectra of radiative and non-radiative shocks on the same plot.

Smith \cite{smith97} catalogues a network of Balmer-dominated filaments surrounding almost the entire remnant and argues that they provide a more complete picture of it than the radiative filaments do.  The morphology of the SNR appears to agree in the H$\alpha$, X-ray (from {\it ROSAT}) and radio (at 843 MHz) images, implying that the non-radiative, optical filaments are a reliable tracer of the blast wave.

Ghavamian et al. \cite{ghava01} measure $W = 562 \pm 18$ km s$^{-1}$ and $I_b/I_n = 1.18 \pm 0.03$ from the southwestern part of the remnant, flanking bright, radiative shocks.  Ghavamian, Laming \& Rakowski \cite{ghava07a} analyze previously unpublished data to add two more data points from the eastern ($W = 640 \pm 35$ km s$^{-1}$, $I_b/I_n = 1.0 \pm 0.2$) and northern ($W = 325 \pm 10$ km s$^{-1}$, $I_b/I_n = 1.06 \pm 0.10$) rims.

Helder et al. \cite{helder09} use {\it Chandra} X-ray proper motion measurements to measure a shock velocity of $v_s = 6000 \pm 2800$ km s$^{-1}$ for northeastern RCW 86.  By contrast, the broad H$\alpha$ line has $W = 1100 \pm 63$ km s$^{-1}$.  If one assumes $v_s \approx W$, then the post-shock temperature is $T_s \approx 2$ keV, in sharp contrast to the temperatures of 42--70 keV derived from the proper motion measurements, suggesting that a substantial fraction of the shock energy is being channelled into cosmic rays (see \S\ref{subsect:cr_injection}).

\subsection{SNRs in the Large Magellanic Cloud}

\subsubsection{SNR 0519--69.0}

Tuohy et al. \cite{tuohy82} are the first to discover BDSs from SNRs outside of our Galaxy.  SNR 0519--69.0, a prominent extended source in a H$\alpha$ image but invisible in [O~{\sc iii}] imaging, is the only remnant in their study for which a reliable detection of the broad H$\alpha$ line is reported, with $W = 2800 \pm 300$ km s$^{-1}$ and $I_b/I_n = 0.4$--0.8.  Smith et al. \cite{smith91} later report $W = 1300 \pm 200$ km s$^{-1}$ from a slower BDS in the same remnant; they also estimate the age of the remnant to be about 500--1500 yrs.

From the intensity of H$\alpha$ and H$\beta$ emission, the density of ambient, neutral hydrogen is estimated by Tuohy et al. to be $\lesssim 0.1$ cm$^{-3}$, in contrast to the total ISM density of about 4.7 cm$^{-3}$ based on analyzing X-ray spectra obtained from the {\it Einstein Observatory}.  This suggests that the ISM around SNR 0519--69.0 is at least 95\% ionized, but such an interpretation rests on the assumptions that the X-ray-emitting plasma has cosmic abundances and is in collisional ionization equilibrium.  If the plasma is enriched by the SN ejecta, then the total ISM density is instead $\gtrsim 0.3$ cm$^{-3}$.

Tuohy et al. also note that SNR 0519--69.0 appears to be a weak radio source at 408 MHz compared to other SNRs in the Large Magellanic Cloud (LMC), implying that either the ISM around it is different or that there exists a fundamental difference in the nature of the particle acceleration process in SNRs with BDSs.  They conclude by speculating that based on the similarity of the observed optical, X-ray and radio properties with the Galactic remnants of Tycho, SN 1006 and Kepler, the progenitor of SNR 0519--69.0 is probably a Type Ia SN.

Ghavamian et al. \cite{ghava07b} use the {\it Far Ultraviolet Spectroscopic Explorer} satellite to measure broad Ly$\beta$ ($W = 3130 \pm 155$ km s$^{-1}$) and O~{\sc vi} $\lambda 1032$ ($W_{\rm O} = 4975 \pm 1830$ km s$^{-1}$) line widths; the detection of bright ($\sim 10$--20\% of Ly$\beta$) O~{\sc vi} emission confirms the prediction of Raymond et al. \cite{ray83}.  The aperture used is comparable to the projected size of the SNR, implying that the broad line widths measured have both intrinsic and and bulk velocity broadening components.  However, the broad H$\alpha$ line width of $2800 \pm 300$ km s$^{-1}$ measured by Tuohy et al. \cite{tuohy82} suggests that the Ly$\beta$ width is dominated by thermal broadening.  In addition to the presence of strong interstellar absorption near the center of the Ly$\beta$ line, the O~{\sc vi} and Ly$\beta$ lines are heavily blended, further adding to the uncertainties in measuring their widths.

\subsubsection{SNR 0505--67.9 (DEM L71)}

SNR 0505--67.9 is discovered in the study of Tuohy et al. \cite{tuohy82}.  Based on its somewhat large size ($\sim 20$ pc, assuming a distance to the LMC of 55 kpc) and the presence of isolated patches of [O~{\sc iii}] detected in imaging, they conclude that it is older than SNRs 0519--69.0 and 0509--67.5.  Like in the case of SNR 0519--69.0, its ambient, neutral hydrogen density is estimated to be $\lesssim 0.1$ cm$^{-3}$ in contrast to the total ISM density of 1 cm$^{-3}$ (based again on assuming cosmic abundances when analyzing X-ray spectra).  It is not detected in the radio at 408 MHz.

Smith et al. \cite{smith91} report the first detection of the broad H$\alpha$ line width ($W = 580 \pm 70$ km s$^{-1}$), but record only an upper limit for the line ratio ($I_b/I_n \ge 0.7$) as they determine the narrow H$\alpha$ line component to be contaminated by radiative shocks.  Interpreting the shock velocity to be 300--800 km s$^{-1}$, they use the Sedov approximation to estimate the age of the remnant to be $2R_{\rm SNR}/5v_s \sim 10^4$ yrs, suggesting that the BDSs are transitioning into radiative shocks.  Smith, Raymond \& Laming \cite{smith94} fail to detect the broad H$\alpha$ emission, which they attribute to poor signal-to-noise in their spectra.

Broad Ly$\beta$ ($W = 1135 \pm 30$, $1365 \pm 75$ km s$^{-1}$) and O~{\sc vi} $\lambda 1032$ ($W_{\rm O} = 740 \pm 45$, $935 \pm 125$ km s$^{-1}$) lines are detected at two locations by Ghavamian et al. \cite{ghava07b}.

\subsubsection{SNR 0509--67.5}

SNR 0509--67.5 is discovered in the study of Tuohy et al. \cite{tuohy82}.  Along with SNR 0519--69.0, it is the only other remnant to be detected at 408 MHz and also appears to be a weak radio source at this wavelength relative to the other LMC remnants.  This is the only remnant associated with BDSs for which a measurement of the broad H$\alpha$ line width has proved to be elusive.

However, broad Ly$\beta$ emission is detected by Ghavamian et al. \cite{ghava07b} --- they measure $W = 3710 \pm 400$ km s$^{-1}$.  Unfortunately, the {\it FUSE} aperture used is essentially the same projected size as the SNR.  The absence of the broad H$\alpha$ line precludes a consistency check for the degree to which the measured broad Ly$\beta$ line width is contaminated by bulk velocity broadening, implying that the measured $W$ value for Ly$\beta$ should be regarded as an upper limit.

\subsubsection{SNR 0548--70.4}

SNR 0548--70.4 is discovered in the study of Tuohy et al. \cite{tuohy82}; imaging reveals patches of isolated [O~{\sc iii}] and the remnant is undetected in the radio at 408 MHz.  The first detection of the broad H$\alpha$ line is reported by Smith et al. ($W = 760 \pm 140$ km s$^{-1}$, $I_b/I_n = 1.1 \pm 0.2$) \cite{smith91}; they also estimate the age of the remnant to be $\sim 10^4$ yrs.

\subsubsection{SNR 1987A}
 
The reverse shock in SNR 1987A may be the highest-velocity BDS ever observed\footnote{In addition to being from a core-collapse supernova.} ($v_s \approx 8000$--9000 km s$^{-1}$) \cite{michael03, heng06}, but such an interpretation --- at least for the ``broad'' H$\alpha$ line --- is controversial.  The neutral hydrogen atoms that produce the BDS emission are not from the ISM, but are instead from the SN ejecta cooled by adiabatic expansion --- they stream across the reverse shock at $\sim 12000$ km s$^{-1}$ in the frame of the observer.  The hydrogen emission with the greatest Doppler shift comes from direct excitation of the atoms, is analogous to the narrow line emission and traces out the surface of the reverse shock --- ``surface emission''.  In the frame of the observer, the broad neutrals produced have lower bulk velocities, are associated with lower Doppler shifts and appear to originate from \emph{beneath} the reverse shock surface --- ``interior emission'' \cite{heng06,heng07b}.  The measured ratio of interior to surface emission is interpreted to be $I_b/I_n \sim 1$.  However, the models of Heng \& McCray \cite{hm07} (see \S\ref{subsect:models}) predict theoretical values of $I_b/I_n \sim 0.1$ for this range of shock velocities, implying that the dominant mechanism responsible for the interior emission may not be associated with a non-radiative shock without a precursor.

\section{Balmer-Dominated Shocks and the Acceleration of Cosmic Rays}
\label{sect:cosmic_rays}

A number of empirical results suggest that the generation of cosmic rays is intrinsic to $v_s \sim 1000$ km s$^{-1}$ BDSs.

\subsection{Narrow H$\alpha$ Lines with Unusually Broad Widths}
\label{subsect:broad_narrow_line}

Smith, Raymond \& Laming \cite{smith94} obtain high resolution spectra of the four LMC remnants studied by Smith et al. \cite{smith91}.  With an instrumental resolution $\sim 13$ km s$^{-1}$, they measure narrow H$\alpha$ line widths $W_0\sim 30$--50 km s$^{-1}$ from shocks at multiple locations in SNRs 0509--67.5, 0519--69.0, 0505--67.9 and 0548--70.4.  Such large widths are difficult to reconcile with the presence of hydrogen emission, which is not expected to exist above $\sim 10^4$ K ($W_0 \sim 20$ km s$^{-1}$; see \S\ref{sect:physics}).  An important clue provided by their observations is that $W_0$ does \emph{not} appear to correlate with $v_s$.  Furthermore, several of the spectra show evidence for non-Gaussian line wings.

Smith, Raymond \& Laming suggest several mechanisms for broadening the narrow H$\alpha$ line\footnote{Precursors are marked with a $\ddagger$ (see \S\ref{sect:physics} for a basic description of precursors).}:
\begin{itemize}

\item Lyman line trapping, in which narrow Ly$\beta$ photons are scattered and converted into H$\alpha$ ones, is a plausible explanation, but it is found to \emph{narrow} the line instead (by $\approx 2$ km s$^{-1}$).

\item Collisions between post-shock protons and pre-shock hydrogen atoms, both prior to and during excitation, may heat the atoms and broaden the narrow line component.  However, both processes are only favorable at high shock velocities ($\gtrsim 3000$ km s$^{-1}$), implying that they cannot account for the $W_0 \gtrsim 30$ km s$^{-1}$ values measured in the older SNRs 0505--67.9 and 0548--70.4.

\item The dissociation of pre-shock molecular hydrogen creates neutral hydrogen atoms with velocities $\sim 30$ km s$^{-1}$, but for $v_s \gtrsim 1000$ km s$^{-1}$ non-dissociative ionization by electrons dominates over all dissociative channels.  Also, H$^+_2$ or H$_2$ are directly converted to fast neutrals via dissociation and charge transfer with protons, respectively, without ever producing slow neutrals.  The $\sim 30$--40 km s$^{-1}$ narrow line widths recorded in the young SNRs 0509--67.5 and 0519--69.0 (which have $v_s \gtrsim 1000$ km s$^{-1}$) make it unlikely that this is the broadening mechanism at work in these remnants.

\item $^\ddagger$If temperature equilibration between the electrons and ions is nearly complete, then the thermal velocity of the electrons exceeds the shock velocity ($v_{{\rm th},e} \sim 23 v_s$; see equation [\ref{eq:vthermal2}]), allowing for thermal conduction via the electrons to heat the pre-shock gas.  Unfortunately, the conduction layer is typically orders of magnitude thinner than the length scale for ionization, implying that the electrons are unable to effectively heat the atoms.  Photoionization precursors suffer from a similar difficulty.

\item $^\ddagger$Magnetohydrodynamic (MHD) precursors rely on a low pre-shock ionization state, such that the Alfv\'{e}n velocity in the ionized gas exceeds the shock velocity upstream.  They generally require ambient magnetic field strengths of a few hundred $\mu$G (e.g., $B_{\rm ISM} \sim 250$ $\mu$G for SNR 0519--69.0).  Furthermore, MHD precursors require a sub-Alfv\'{e}nic but still supersonic shock --- the two components will have different shock structures in general (see \S3.2 of Draine \& McKee \cite{dm93}).

\item $^\ddagger$An interesting possibility that has not been fully explored (see \S\ref{subsect:broad_neutral_precursor}) is that a substantial fraction of the broad neutrals, which flow away from the shock front at $\sim v_s/4$, may have thermal velocities exceeding the flow velocity.  They are thus capable of imparting energy to the pre-shock gas.  This process is expected to be sensitively dependent upon the shock velocity, which makes it potentially at odds\footnote{The velocity-dependence of precursor heating by broad neutrals has not been clearly demonstrated (\S\ref{subsect:broad_neutral_precursor}).} with the empirical finding that $W_0$ appears to be uncorrelated with $v_s$.

\item $^\ddagger$An intriguing possibility is that energetic particles scattering between the shock front and a pre-shock, turbulent region (i.e., cosmic rays) will heat the pre-shock gas (see \S\ref{subsect:cosmic_ray_precursor}).  In perpendicular shocks --- where the magnetic field lines are perpendicular to the flow direction --- if the broadening is associated with the conservation of the adiabatic invariant \cite{rakow09},
\begin{equation}
\frac{v^2_\perp}{B_{\rm ISM}} = \mbox{constant},
\end{equation}
where $v_\perp \sim W_0$ is the velocity component of an electron or ion perpendicular to the field line, then a factor $\sim 3$ increase in $W_0$ corresponds to about an order of magnitude amplification of the ambient magnetic field.  In contrast to the other suggested mechanisms, the concern here is that the heating is \emph{too} efficient and the precursor thickness should be small enough that the pre-shock gas is not substantially ionized.  The upper limit to the cosmic ray diffusion coefficient (equation [\ref{eq:cr_coeff}]) is orders of magnitude below typical values for the undisturbed ISM, but this can be reconciled with the fact that the pre-shock region of a BDS is partially neutral --- ion-neutral collisions are expected to heavily damp the Alfv\'{e}n waves \cite{drury96}, diminish the effective value of $\kappa_{\rm CR}$, and limit the fraction of shock energy deposited into cosmic rays.  It is not clear if heating by a cosmic ray precursor has a strong dependence on the shock velocity.

\end{itemize}

Hester, Raymond \& Blair \cite{hester94} measure $W_0 \sim 33$ km s$^{-1}$ at multiple locations in the Cygnus Loop, which has the lowest shock velocities ($v_s \lesssim 200$ km s$^{-1}$) among SNRs with BDSs.  Their kinetic models show that radiative transfer effects are unable to account for this somewhat large width, consistent with the conclusions of Smith, Raymond \& Laming.

Ghavamian et al. \cite{ghava00} record $W_0 = 44 \pm 4$ km s$^{-1}$ from high-resolution spectra of knot g in Tycho's remnant and also detect diffuse H$\alpha$ emission extending $\gtrsim 1$ pc ahead of the knot.  They attribute the diffuse emission to a photoionization precursor.  Since non-radiative shocks lack recombination zones, they reason that the main source of photoionization is photons from lines in the immediate post-shock region.  Models in which He~{\sc ii} $\lambda 304$ is the primary source of ionizing radiation from behind the forward shock match observed [S~{\sc ii}]/H$\alpha$ and [O~{\sc iii}]/H$\beta$ --- but not [N~{\sc ii}]/H$\alpha$ --- line ratios fairly well, but fall short of explaining the observed narrow H$\alpha$ line width ($T_0 \approx 4\times 10^4$ K).  They speculate that a second precursor might be present.  Curiously, the narrow H$\alpha$ line shows weak evidence for the presence of a second, broader component --- fitting a second Gaussian to the line yields a broader component with width $\sim 150$ km s$^{-1}$ --- the origin of which is unexplained.

Sollerman et al. \cite{sollerman03} use high resolution spectroscopy to measure narrow H$\alpha$ line widths in RCW 86 and the remnants of Kepler and SN 1006.  Their Figure 6 demonstrates that the correlation between $W_0$ and $v_s$ is not straightforward --- the narrow line width remains somewhat constant for most of their employed data points ($W_0 \sim 30$--40 km s$^{-1}$), while their highest-velocity data point (for SN 1006) has $W_0 = 21 \pm 3$ km s$^{-1}$.  Curiously, SNR 0509--67.5 has $W_0 = 25$--31 km s$^{-1}$ but no published measurement of the broad H$\alpha$ line width, although Tuohy et al. \cite{tuohy82} estimate $v_s \gtrsim 3600$ km s$^{-1}$ for this remnant.  Thus, there is tentative evidence for a cut-off in the narrow H$\alpha$ line width at high shock velocities, the origin of which is not understood.

\subsection{Unusually Low Broad-to-Narrow H$\alpha$ Line Intensity Ratios}
\label{subsect:large_ibin}

DEM L71 --- otherwise known as SNR 0505--67.9 --- is the poster child for demonstrating that BDS models without precursors (\S\ref{subsect:models}) fail to capture essential physics.  Ghavamian et al. \cite{ghava03} measure 14 values for $W$ and $I_b/I_n$  around the entire periphery of the remnant.  While the broad H$\alpha$ FWHM is $\lesssim 1000$ km s$^{-1}$, the mean value of $I_b/I_n$ is about 0.5; in one location, $I_b/I_n$ is as low as 0.2.  Most of the measured $I_b/I_n$ values are too low to be explained by either the models of Ghavamian et al. or by the models of Heng \& McCray \cite{hm07} and van Adelsberg et al. \cite{v08} (see \S\ref{subsect:models}).

In addition to broadening the narrow line, precursors contribute to the narrow line intensity, hence lowering the value of $I_b/I_n$, conceivably explaining the discrepancy between models and observations and motivating the need for improved models that include cosmic ray precursors (see \S\ref{subsect:models_precursors}).  Ghavamian et al. \cite{ghava03} find no evidence for diffuse H$\alpha$ emission ahead of the BDSs, implying that the precursor thickness is comparable to or less than the thickness of the observed filaments.

Rakowski, Ghavamian \& Laming \cite{rakow09} confirm the low $I_b/I_n$ values obtained by Ghavamian et al. with higher-resolution spectroscopy, reporting measured values of $I_b/I_n$ at  16 positions, again with a mean value of about 0.5.

\subsection{Acceleration of Cosmic Rays in Partially Neutral Media}
\label{subsect:pneutral}

\subsubsection{The Remnant of SN 1006}

It is previously known that the remnant of SN 1006 is ``bipolar'' or barrel-shaped, being brightest (in the radio) in the northeast and southwest.  Early X-ray spectra of SN 1006 are entirely featureless, hinting at a non-thermal (i.e., synchrotron) origin.  Coupled with the absence of evidence for a central pulsar, this led to speculation that the non-thermal X-rays originate from electrons accelerated at the shock front by the first-order Fermi mechanism.

In an insightful study, Winkler \& Long \cite{wl97} compare H$\alpha$, X-ray ({\it ROSAT}) and radio (1370 MHz) maps of SN 1006.  Predominantly thermal X-ray emission is found around the entire periphery of SN 1006, spatially offset from the H$\alpha$ filaments and presumably tracing out the post-shock cooling region.  In the northeast and southwest, the X-ray spectrum is better described by a power law; the detailed correspondence between the X-ray and radio structures support a common synchrotron origin for emission in both limbs.

If the ambient magnetic fields are amplified to $\sim 100$ $\mu$G, then radio synchrotron emission at $\sim$ GHz frequencies is associated with $E_e \sim 1$ GeV cosmic ray electrons,
\begin{equation}
\nu_{\rm radio} \approx 0.5 \mbox{ GHz} ~\left(\frac{E_e}{1 \mbox{ GeV}} \right)^2 \left(\frac{B_{\rm ISM}}{100 ~\mu\mbox{G}} \right),
\end{equation}
while X-ray synchrotron emission at $\sim$ keV energies originates from $\sim 10$ TeV cosmic ray electrons,
\begin{equation}
E_{\rm X-ray} \approx 3.0 \mbox{ keV} ~\left(\frac{E_e}{30 \mbox{ TeV}} \right)^2 \left(\frac{B_{\rm ISM}}{100 ~\mu\mbox{G}} \right).
\end{equation}
Cosmic ray protons at $\sim$ keV energies are expected to produce non-radiative optical emission (see \S\ref{subsect:pickup}).  A review of particle acceleration in SNRs can be found in Reynolds \cite{reynolds08}.

\begin{figure}
\begin{center}
\includegraphics[width=\columnwidth]{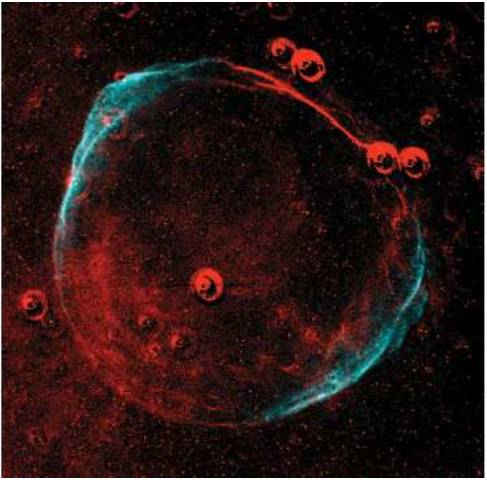}
\end{center}
\vspace{-0.2in}
\caption{Combined optical and X-ray image of the remnant of SN 1006, adapted from Cassam-Chena\"{i} et al. \cite{cc08}.  Shown are the {\it Chandra} synchrotron (2--4.5 keV; in cyan) \cite{cc08} and {\it CTIO Curtis Schmidt} narrow-band H$\alpha$ (in red) \cite{winkler03} emission.  Courtesy of Gamil Cassam-Chena\"{i} and Jack Hughes.  (North is up and east is left.)}
\label{fig:sn1006}
\end{figure}

Optical, H$\alpha$ filaments are detected around much of the periphery of SN 1006, but are most prominent in the density-enhanced northwest region, which has neutral densities $\sim 0.1$--1 cm$^{-3}$ compared to $\sim 0.01$--0.1 cm$^{-3}$ for the rest of the remnant.  A sharply-defined radio shock is not seen at the northwest limb; together with the lack of correlation with the X-ray emission, this suggests the synchrotron component to the emission is negligible.  This picture is corroborated by the fact that the spatial variation of the X-ray surface brightness behind the shock front is adequately modelled by a cooling thermal plasma.  The spatial offset between the peaks of the X-ray and H$\alpha$ intensities implies a temporal delay of $\sim 120$ yrs between the ionization of neutral species and the production of thermal X-rays; the radio emission develops $\sim 600$ yrs after ionization.

Cassam-Chena\"{i} et al. \cite{cc08} re-investigate the optical, X-ray and radio properties of SN 1006, focusing on the poorly-explored southeast limb where a somewhat fainter --- but still clearly-defined --- H$\alpha$ filament can be seen.  Again, synchrotron X-ray and radio emission appear to be suppressed.  However, they also find that over small ranges near the eastern and southern azimuths, synchrotron and non-radiative H$\alpha$ emission appear to be co-spatial (see Figure \ref{fig:sn1006}), making SN 1006 one of the rare cases where such a correlation has been observed.

Using the BDS in H$\alpha$ and thermal X-rays as respective tracers of the blast wave (BW) and contact discontinuity (CD) in SN 1006, Cassam-Chena\"{i} et al. are able to quantify the ratio of the BW to CD radii, $R_{\rm BW}/R_{\rm CD}$.  They record $R_{\rm BW}/R_{\rm CD} \approx 1$ in places where synchrotron emission is present; near the southeast azimuth, where synchrotron emission is suppressed, they measure $R_{\rm BW}/R_{\rm CD} \approx 1.1$.  These empirical values are consistent with the notion that the production of cosmic rays reduces the pressure between the BW and CD, and allows them to be located closer to each other (as Warren et al. \cite{warren05} find for Tycho's remnant).  Also, it is found that the brighter the BW, the smaller the separation between the BW and CD.  Using cosmic ray-modified hydrodynamic models, Cassam-Chena\"{i} et al. are able to account for the azimuthal variation of the X-ray and radio brightnesses by allowing the injection rate of particles, ambient magnetic field and cosmic ray diffusion coefficient to vary azimuthally.  A major shortcoming of their model is the inability to produce $R_{\rm BW}/R_{\rm CD} \le 1.1$ values; an ad hoc reduction factor of 12\% is applied to all of their results on the assumption that it is caused by an effect unrelated to the cosmic ray acceleration process.

\subsubsection{Tycho's Remnant}

The presence of thin, bright, X-ray synchrotron-emitting rims --- identified in the {\it Chandra} 4.1--6.1 keV band\footnote{Warren et al. estimate the thermal contribution to the 4.1--6.1 keV band at the rims to be $\sim 10\%$.} by Warren et al. \cite{warren05} --- at the BW all around Tycho's remnant implies that there exist locations where BDS and non-thermal emission are correlated, although these regions have not yet been scrutinized to the same level as for the remnant of SN 1006.  Warren et al. conclude that the mere \emph{existence} of the BDSs argues for the efficiency of cosmic ray acceleration to vary around the rim of Tycho's SNR.

\section{Theoretical Models}
\label{sect:theory}

Theoretical models of BDSs can be broadly categorized into those without (\S\ref{subsect:models}) and with (\S\ref{subsect:models_precursors}) precursors.  There is a general disconnect between the models in \S\ref{subsect:models} and \S\ref{subsect:models_precursors} --- the former compute detailed, broad hydrogen line profiles and $I_b/I_n$ values while ignoring physics that broaden the narrow hydrogen line, while the latter isolate precursor physics without unifying them with the non-precursor models in a self-consistent manner.

\subsection{Models Without Precursors}
\label{subsect:models}

Observations of BDSs from SNRs commonly yield $I_b/I_n < 1$, which early models are unable to account for.  For example, the models of Smith et al. \cite{smith91} predict $I_b/I_n > 1$, due to various simplifications in the way they treat the atomic interactions.

Ghavamian et al. \cite{ghava01} improve upon the model of Chevalier, Kirshner \& Raymond \cite{ckr80} by including an approximate treatment of the shock transition zone --- the region $\sim 10^{15}(1\mbox{ cm}^{-3}/n)$ cm where excitation, ionization and charge transfer take place --- and a Monte Carlo calculation of Lyman line trapping (see \ref{subsect:atomic0}) in their models.  The latter involves following the propagation, absorption and conversion of Lyman photons into Balmer ones in the pre-shock region.  Electron-proton temperature equilibration is parametrized by
\begin{equation}
f_{\rm eq} \approx \frac{2\beta}{1+\beta}.
\label{eq:feq}
\end{equation}

To compare models with observations (of the Cygnus Loop, Tycho's SNR and RCW 86), Ghavamian et al. compute $I_b/I_n$ versus $0 \le f_{\rm eq} \le 1$ for different values of the pre-shock ion fraction $f_{\rm ion}$.  These calculations are then compared to the observed ranges of $I_b/I_n$ to constrain the range of allowed $f_{\rm eq}$ values.  The dependence on $f_{\rm ion}$ is found to be somewhat weak, but in the case of southwestern RCW 86 does allow $f_{\rm ion} \gtrsim 0.9$ to be ruled out.  In the case of Tycho's knot g, the Ghavamian et al. models are unable to produce the observed $I_b/I_n = 0.67 \pm 0.1$ value for H$\alpha$; the theoretical predictions are $\gtrsim 1$.

In the Ghavamian et al. model, the effects of multiple charge transfers on the broad neutral distribution are not treated.  The pre-shock temperature $T_0$ is a free parameter that is set to 5000 K.  This technique of constraining the degree of electron-proton equilibration is similar to that employed later by van Adelsberg et al. \cite{v08} (see equation [\ref{eq:equilibration}]).

Heng \& McCray \cite{hm07} revisit the theory of BDSs and present a formalism that allows one to compute the broad neutral velocity distributions, rate coefficients and probabilities for atomic interactions from first principles, with the only input being the atomic cross sections.  Multiple reactions involving the same atom are visualized using a ``reaction tree'' (their Figure 4), allowing for multiple charge transfers and excitations to be included in computing the rate coefficients.  Consequently, the Heng \& McCray model predicts theoretical values for $W$ and $I_b/I_n$ as a function of $v_s$ and $\beta$; only the $\beta = 0.25$ and $\beta=1$ models are published.  The narrow hydrogen line is approximated by a Dirac-delta function in velocity space, i.e., $T_0 = 0$.  The velocity offset between the centers of the narrow and broad distribution profiles yields the inclination of the plane-parallel shock relative to the line of sight, e.g., $\theta \approx 6^\circ$ for the Kirshner, Winkler \& Chevalier \cite{kwc87} observation of the nearly edge-on BDS in Tycho's remnant (see \S\ref{subsect:tycho}).

Heng et al. \cite{heng07} include the effects of $f_{\rm ion}$ and $T_0$ into the Heng \& McCray model by treating the shock transition zone in the approximation that the broad neutrals and post-shock ions share a common bulk velocity and temperature.  The expectation that the Rankine-Hugoniot jump conditions are obeyed at the shock front comes naturally out of the models.  The models also allow the spatial emissivities of the narrow and broad H$\alpha$ lines to be calculated, providing a consistency check\footnote{It is worth noting that for the purpose of predicting the spatial shift between the narrow and broad line emissivities --- at least for SN 1006 --- it is sufficient to use the two-component model of Heng et al., which is analytical.} on the Raymond et al. \cite{ray07} computations for SN 1006.  Lyman line trapping is not treated in the Heng et al. model.

Van Adelsberg et al. \cite{v08} generalize the Heng et al. model to fully incorporate the effects of multiple charge transfers on the broad neutral population and relax the assumption that broad neutrals and post-shock ions share a common bulk velocity.  Lyman line trapping is treated using an approximate fitting function.  The more rigorous treatment of the shock transition zone produces lower theoretical values of $I_b/I_n$, thus allowing self-consistent values of $v_s$ and $\beta$ to be derived from measured values of $W$ and $I_b/I_n$ (see \S\ref{subsect:equilibration}).  For example, for the $W=1765 \pm 110$ km s$^{-1}$, $I_b/I_n = 0.67 \pm 0.1$ values observed in Tycho's knot g \cite{ghava01}, van Adelsberg et al. obtain $v_s \approx 1600$ km s$^{-1}$ and $\beta \approx 0.05$.

\subsection{Models With Precursors}
\label{subsect:models_precursors}

\subsubsection{Cosmic Ray Precursor}
\label{subsect:cosmic_ray_precursor}

Rakowski, Laming \& Ghavamian \cite{rakow08} study electron heating by lower hybrid waves, which are electrostatic ion waves propagating nearly perpendicular to the magnetic field lines.  They have a frequency equal to the geometric mean of the electron and ion gyrofrequencies.  The growth of lower hybrid waves is most efficient in quasi-perpendicular shocks, i.e., the magnetic field is nearly perpendicular to the flow across the plane-parallel shock.  Lower hybrid waves have the property that their phase velocity parallel to the field lines greatly exceeds its perpendicular counterpart, thereby allowing collisionless energy exchange to occur between ions moving across --- and electrons moving along --- the lines.

Above the critical Alfv\'{e}nic Mach number of ${\cal M}_{A,{\rm crit}} \sim 12$--60, cosmic ray ions channel their energy into amplifying the magnetic field; even an initially quasi-parallel shock evolves into a quasi-perpendicular one via the development of a helical magnetic field.  Below ${\cal M}_{A,{\rm crit}}$, the dominant mechanism for the cosmic ray ions to deposit energy is through the generation of lower hybrid waves, which accelerate a small fraction of electrons that are in resonance.  These accelerated electrons communicate their energy to the rest of the thermal population via Coulomb interactions.  Magnetic field amplification and lower hybrid waves work together --- as the field close to a high Mach number-shock front gets amplified, the effective value of ${\cal M}_A$ is reduced, eventually allowing lower hybrid wave growth to take over.  Heating is accomplished via damping of the waves.  Consequently, there are two precursors: an extended ($\sim 10^{18}$ cm) precursor in which amplification occurs, followed by a narrower ($\sim 10^{15}$ cm) precursor in which electron heating happens.

A key assumption of the Rakowski, Laming \& Ghavamian model is that cosmic ray ions are ubiquitous in SNR shocks.  If their hypothesis regarding lower hybrid waves is correct, it implies that cosmic rays are not by-products of collisionless shocks --- rather, they are an intrinsic component that affects the shock structure, dynamics and energetics.

Wagner et al. \cite{wagner09} use a two-fluid approximation to model the back-reaction of the cosmic rays on the thermal, pre-shock gas --- the cosmic rays are treated as a massless fluid exerting a bulk pressure; their particle distribution is not followed in the calculation.  Heating is accomplished by the damping of sound waves emitted by the cosmic rays via an acoustic instability.  Shock parameters appropriate to knot g in Tycho's SNR are adopted: $v_s = 2000$ km s$^{-1}$, $T_0 = 1.2 \times 10^4$ K, $n = 1$ cm$^{-3}$ and $\kappa_{\rm CR} = 2 \times 10^{24}$ cm$^2$ s$^{-1}$.  Cosmic ray injection at the gas sub-shock\footnote{In cosmic ray physics, the fluid layer where plasma heating occurs is termed the ``gas sub-shock''.} is included in the model and shown to be necessary in order to match the observed spatial emissivities of narrow and broad H$\alpha$ emission.  However, matching the spatial emissivities also requires a pre-sub-shock temperature of $10^5$ K, somewhat higher than the $T_0 \approx 4 \times 10^4$ K value implied by the Lee et al. \cite{lee07} observations of Tycho's knot g if interpreted within the context of a cosmic ray precursor.

Initially, the shock is weakly modified\footnote{Meaning that the adiabatic index of the gas is $\gamma \approx 5/3$ instead of 4/3.} by cosmic rays, but rapidly ($\sim 1500$ yrs) evolves into a cosmic ray-dominated shock.  A  transient solution for the shock that satisfies the observational constraints develops  within 420 yrs after the initial state.  About 1000 yrs post-initialization, a density spike develops in the solution due to a temporary over-compression associated with both the gas sub-shock and the cosmic ray precursor.  Wagner et al. predict that the density spike may manifest itself as a factor $\sim 2$--4 enhancement in the thermal X-ray intensity.  Furthermore, they discover a branch of steady solutions with low cosmic ray acceleration efficiency --- which they do not match to the observed spatial emissivities --- that are relatively insensitive to the shock parameters.  Wagner et al. speculate that these solutions may account for the lack of correlation between $W_0$ and $v_s$ (see \S\ref{subsect:broad_narrow_line}).

\subsubsection{Broad Neutral Precursor}
\label{subsect:broad_neutral_precursor}

Lim \& Raga \cite{lr96} investigate the broadening of the narrow hydrogen line via elastic collisions with broad neutrals.  Their study considers a range of pre-shock ionization fractions, since it is likely the main parameter that controls the upstream energy flux, but only for face-on shocks with $v_s = 225$ km s$^{-1}$.  However, they find a nearly negligible dependence: for $f_{\rm ion} = 0.5$--0.99, $W_0$ decreases from about 22.6 to 21.6 km s$^{-1}$.

The atomic cross sections and rate coefficients employed in the calculations of Lim \& Raga are from references in or before 1970 (see \S\ref{subsect:atomic}).  The key process studied, ion-neutral elastic scattering, is modelled using an approximate rate coefficient.  A Gaussian distribution is assumed for the broad neutrals --- while this is a good approximation at $v_s = 225$ km s$^{-1}$, it breaks down at high shock velocities because charge transfer becomes inefficient compared to ionization (see \S\ref{subsect:atomic}).  It remains to be shown how sensitively the increase in $W_0$ depends on $v_s$.

\subsection{The Relevance of Atomic Physics}
\label{subsect:atomic}

\subsubsection{Charge Transfer, Excitation and Ionization}
\label{subsect:atomic0}

The collisionless, non-radiative shock model \cite{cr78,ckr80,hm07} predicts that cold, pre-shock neutral atoms stream past the shock front without acceleration or ionization, and subsequently engage in charge transfer/exchange, excitation or ionization reactions with post-shock electrons and ions.  The cross section for ionization exceeds that for charge transfer to the ground state at just under $\Delta v = 3000$ km s$^{-1}$ (e.g., see Figure 1 of \cite{hm07}), somewhat coincident with the thermal velocity of an electron with 13.6 eV of energy, $v_{{\rm th},e} \approx 2700$ km s$^{-1}$ (equation [\ref{eq:vthermal}]).

The dominance of charge transfer/exchange for $\Delta v \lesssim 3000$ km s$^{-1}$ ($v_s \lesssim 4000$ km s$^{-1}$) implies that the broad H$\alpha$ line profile and \emph{projected}\footnote{Meaning for a given viewing angle between the shock normal and the line of sight.} post-shock proton distribution are virtually identical.  In this regime, we regard charge transfer as being ``efficient''.  The mapping becomes more complicated at higher velocities because excitation and ionization competes effectively with charge transfer.

Direct collisional excitation of the pre-shock atoms,
\begin{equation}
\mbox{H} + \{e^-, p\} \rightarrow \mbox{H}^\ast + \{e^-, p\},
\end{equation}
results in narrow hydrogen emission lines with widths that reflect the thermal conditions of the pre-shock gas ($\sim 10$ km s$^{-1}$; but see \S\ref{sect:cosmic_rays}).  For $v_s \lesssim 2000$ km s$^{-1}$, the rate coefficients are dominated by electron impact excitation; for $v_s \gtrsim 2000$ km s$^{-1}$, excitation by protons takes over (see Figure 3 of \cite{hm07}).  Singly- and doubly-ionized helium ions may also excite the atoms, but this is a secondary effect unless the abundance of helium is appreciably more than 10\%.

A secondary population of hydrogen atoms, i.e., the broad neutrals, is created when the pre-shock hydrogen atoms exchange electrons with post-shock protons,
\begin{equation}
\mbox{H} + p \rightarrow
\begin{cases}
p + \tilde{\mbox{H}}\\
p + \tilde{\mbox{H}}^\ast\\
\end{cases}
,
\end{equation}
where the broad neutrals are labelled as $\tilde{\mbox{H}}$.  Charge transfers to both the ground and excited states have to be accounted for in any reasonable model \cite{ckr80,ghava01,hm07}.

Experimentally, one can distinguish between elastic scattering and charge transfer at high energies because the fast emerging particle is either a proton or a hydrogen atom \cite{schultz08}.  However, quantum indistinguishability prevents such a distinction at very low velocities ($\ll 1$ eV); the quantum spin is then the only way to label the proton.  Consequently, charge transfer at very low energies is termed ``spin exchange''.

\emph{It is best to think of BDS emission as a probabilistic event --- a given hydrogen atom may engage in charge transfer, excitation or ionization at any moment \cite{hm07}.  Its ultimate fate is always ionization; the question is how many Balmer, Lyman, Paschen, etc., photons will it emit before perishing?}

The BDSs that have been observed so far occur in a velocity range that is interesting for atomic physics.  The relative velocity between interacting atoms and ions, $\Delta v \sim 3v_s/4 \sim 1000$ km s$^{-1}$, is in the range where neither the processes of charge transfer, excitation nor ionization can be ignored.

If the pre-shock gas is optically thin to Lyman lines, it has Case A conditions.  If it is optically thick, then Case B conditions are present.  For example, for Case A about 88\% of excitations to the $3p$ level result in Ly$\beta$ photons ($3p \rightarrow 1s$), while the remaining 12\% yield H$\alpha$ photons ($3p \rightarrow 2s$); if Case B conditions prevail, all of the Ly$\beta$ photons are converted into H$\alpha$ ones via scattering, a phenomenon known as ``Lyman line trapping''.  Often, the intermediate regime occurs in BDSs, because in the case of the narrow Ly$\beta$ and Ly$\gamma$ lines the optical depths are \cite{ckr80}:
\begin{equation}
\begin{split}
&\tau\left(\mbox{Ly}\beta\right) \approx  \frac{4 \delta \left(1-f_{\rm ion}\right)}{f_{\rm ion}} \left(\frac{T_0}{10^4 \mbox{ K}} \right)^{-1/2},\\
&\tau\left(\mbox{Ly}\gamma\right) \approx  \frac{1.4 \delta \left(1-f_{\rm ion}\right)}{f_{\rm ion}} \left(\frac{T_0}{10^4 \mbox{ K}} \right)^{-1/2},
\end{split}
\end{equation}
where $f_{\rm ion}$ is the pre-shock ion fraction and $\delta$ is fraction of neutral gas at a temperature $T_0$.  The larger number of scatterings experienced by Ly$\beta$ photons results in a more complete conversion of Ly$\beta$ to H$\alpha$ than of Ly$\gamma$ to H$\beta$, thus enhancing the relative intensity of H$\alpha$.  As such, the ``Balmer decrement'', defined as the intensity ratio of H$\alpha$/H$\beta$, generally increases with optical depth.  High values of the Balmer decrement may indicate the increased importance of collisions, but this situation is degenerate with the presence of dust, which preferentially attenuates H$\beta$ photons.

\subsubsection{References for Atomic Cross Sections}

For convenience, we summarize the most modern list of references by atomic process:
\begin{itemize}

\item Charge transfer/exchange: Janev \& Smith \cite{janev93}; Belki\'{c} et al. \cite{belkic92}; Harel, Jouin \& Pons \cite{harel98}, Schultz et al. \cite{schultz08}.

\item Excitation: Janev \& Smith \cite{janev93}; Bubelev et al. \cite{bubelev95}; Balan\c{c}a, Lin \& Feautrier \cite{balanca98}.

\item Ionization: Janev \& Smith \cite{janev93}.

\end{itemize}

For processes involving neutral helium, singly-ionized helium and alpha particles:
\begin{itemize}

\item Charge transfer/exchange: Barnett et al. \cite{barnett90}; Janev \& Smith \cite{janev93}; Hose \cite{hose95}.

\item Excitation: Janev \& Smith \cite{janev93}; Bray et al. \cite{bray93}.

\item Ionization: Peart, Walton \& Dolder \cite{peart69}; Angel et al. \cite{angel78}; Rudd et al. \cite{rudd83}; Shah \& Gilbody \cite{shah85}; Rinn et al. \cite{rinn86}; Shah et al. \cite{shah88}; Janev \& Smith \cite{janev93}.

\end{itemize}

Other references may be found in Laming et al. \cite{laming96}.  Several fitting functions for some of the atomic data (in the references listed above) are given in the appendices of Heng \& Sunyaev \cite{heng08} and van Adelsberg et al. \cite{v08}.  We note that Schultz et al. computed cross sections for proton-hydrogen atom charge transfer (and spin exchange) for $200 \lesssim v_s \lesssim 20000$ km s$^{-1}$ using a variety of techniques.

\subsubsection{The Scarcity of Atomic Cross Sections}
\label{subsect:scarcity}

Even with modern techniques, the uncertainties in the computed cross sections can be $\sim 10$--30\%, representing a non-negligible source of systematic error for any model of BDSs that incorporates the effects of atomic physics.  In particular, the proton-hydrogen atom collision system remains an active area of research \cite{fritsch91,martin99,schultz08,fathi09}.  Cross sections for the excitation of hydrogen atoms by protons at the velocities relevant for BDSs remain scarce.

The proton-hydrogen atom collision system is qualitatively described by three different energy regimes with no clear boundaries between them \cite{fritsch91}.  In the low-energy regime, charge transfer is the dominant, resonant channel, since almost no energy needs to be transferred from the relative nuclear to the electronic motion --- the system may be regarded as a quasi-molecule with a temporally-varying internuclear separation \cite{schultz08}.  In the high-energy regime, charge transfer becomes a very weak process compared to excitation and ionization --- the projectile then provides a small perturbation to the wavefunction of the target.  In the intermediate-energy regime, the interplay between charge transfer, excitation and ionization is important; in this regime, obtaining detailed results like the population of projectile or target sub-shells is computationally challenging.

Roughly speaking, for projectiles and targets of nearly equal mass, atomic cross sections are somewhat difficult to compute when
\begin{equation}
\frac{\Delta v}{\alpha_{\rm FS} c} \sim 1 ~\Rightarrow ~\frac{m_p \Delta v^2}{2} \sim 25 \mbox{ keV},
\end{equation}
where $\alpha_{\rm FS} \approx 1/137$ is the fine structure constant.  For example, the one-center, close-coupling\footnote{The adjective ``coupling'' comes from having to solve a pair of coupled Schr\"{o}dinger equations.} expansion \cite{martin99} commonly used to solve the time-dependent Schr\"{o}dinger equation, in order to obtain proton-hydrogen atom excitation cross sections, is inaccurate below impact energies $\sim 20$--30 keV, because charge transfer becomes the dominant channel and computational convergence is thus difficult to achieve.  Above energies of a few hundred keV, perturbative methods may be employed \cite{fritsch91}.

A crude scaling law to obtain excitation --- or charge transfer to excited states --- cross sections to the level $n_2 l_2$ is
\begin{equation}
\sigma_{n_2 l_2} = \left(\frac{n_1}{n_2} \right)^3 \sigma_{n_1 l_1},
\end{equation}
where $\sigma_{n_1 l_1}$ is the cross section to the atomic level $n_1 l_1$ ($n_1<n_2$).  This relation should only be used for order-of-magnitude computations or as a last resort in the absence of available atomic data.

The glaring lack of cross sections for the excitation of hydrogen atoms by protons to the levels $4s$, $4p$, $4d$, $4f$ and above, at $\Delta v \sim 1000$ km s$^{-1}$, implies that \emph{Balmer decrements cannot be computed accurately enough for them to be used as shock diagnostics at $v_s \sim 1000$ km s$^{-1}$}.

\section{Balmer-Dominated Shocks in Other Astrophysical Objects}
\label{sect:others}

The abundance of hydrogen present in the universe and the existence of BDSs across a wide range of velocities suggest that they should be present in other astrophysical objects as well.  In this section, I comment on the implications of both probable and possible detections of BDSs.  A basic but key test is to demonstrate that the detected hydrogen line can account for the measured luminosity assuming plausible values for the physical parameters, despite originating from a non-radiative shock.

\subsection{Pulsar Wind Nebulae}

\begin{figure}
\begin{center}
\includegraphics[width=\columnwidth]{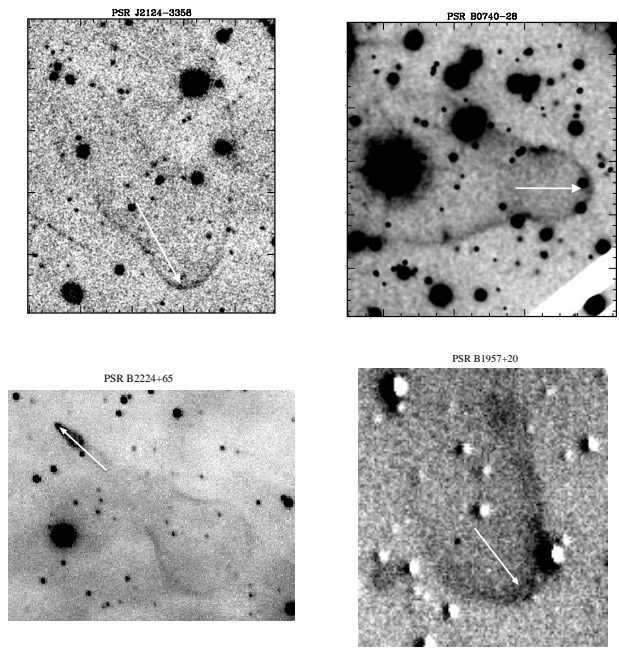}
\end{center}
\vspace{-0.2in}
\caption{Montage of narrow-band H$\alpha$ images of the optical bow shocks associated with pulsar wind nebulae.  On display are J2124--3358 \cite{gaensler02}, B0740--28 \cite{jones02}, B1957 +20 \cite{stappers03} and B2224 +65 \cite{cc02}.  The arrows mark the positions of the pulsars.  Courtesy of Bryan Gaensler.  (North is up and east is left.)}
\label{fig:pwn}
\end{figure}

Old ($\gtrsim 10^9$ yrs), milli-second pulsars --- neutron stars that are spun up by the accretion of material from a companion --- lose their rotational energy by driving a wind into the ISM.  One naturally expects the wind to manifest itself as a BDS in H$\alpha$, but such detections are rare\footnote{Cordes, Romani \& Lundgren \cite{cordes93} estimate that $\sim 5000$ Galactic pulsars may be able to excite H$\alpha$ nebulae.} because the ISM around these pulsar wind nebulae is expected to be either fully-ionized or of low density.  Nevertheless, several examples have been found (Figure \ref{fig:pwn}), spanning a range of distances ($\sim 0.1$--2 kpc) and spin-down luminosities ($\sim 10^{33}$--$10^{35}$ erg s$^{-1}$).

Kulkarni \& Hester \cite{kh88} discover, via H$\alpha$ imaging, a distinct, comet-shaped nebula around the pulsar PSR B1957 +20, which has a period of 1.6 ms.  The pulsar appears to be moving through the ISM at $\sim 100$ km s$^{-1}$, producing a shock of comparable velocity.  About 1--10\% of the rotational luminosity (of about $3 \times 10^{35}$ erg s$^{-1}$) is channelled into H$\alpha$ emission.  The spectrum reveals weak or absent [O~{\sc iii}] and [S~{\sc ii}] emission alongside a prominent, narrow H$\alpha$ line, consistent with the properties of a BDS.  The broad H$\alpha$ line is not detected.  The narrow H$\beta$ line is barely seen --- assuming a positive detection, the dust-corrected Balmer decrement is $\gtrsim 6$, at least twice as large as the expected value of 3.

With improved spectroscopy, Aldcroft, Romani \& Cordes \cite{aldcroft92} clearly detect H$\alpha$ and H$\beta$ lines from the bow shock around PSR B1957 +20; the nebular lines of heavy elements have upper limits of 3.5\% of the H$\alpha$ line intensity.  The mean Balmer decrement is determined to be H$\alpha$/H$\beta = 3.5 \pm 0.5$, implying that the visual extinction is $A_V \lesssim 1$.  A single Gaussian fit to the H$\alpha$ line indicates that the face-on, bulk velocity of the flow is $\sim 120$ km s$^{-1}$.  A five-parameter kinetic model is fitted to the observed spectrum: the ambient density is $0.41 \pm 0.02$ cm$^{-3}$ and the distance to the pulsar is $1.2 \pm 0.01$ kpc.  Aldcroft, Romani \& Cordes remark that, at least in the case of PSR B1957 +20, the analysis of the H$\alpha$ line provides the cleanest measurement of the distance, as noise in the radio pulsar and its faintness render conventional parallex methods unlikely to succeed.

Following the discovery of the optical nebula around PSR B1957 +20, several other examples are found.  Cordes, Romani \& Lundgren \cite{cordes93} present a Balmer-dominated spectrum of the Guitar nebula produced by the high-velocity ($\gtrsim 800$ km s$^{-1}$) pulsar PSR B2224 +65.  Bell et al. \cite{bell95} report an image of the H$\alpha$ bow shock around PSR J0437--4715.  Van Kerkwijk \& Kulkarni \cite{vk01} present spectra and H$\alpha$ images of the nebula around the neutron star RX J1856.5--3754.  Jones, Stappers \& Gaensler \cite{jones02} image an optical bow shock associated with PSR B0740--28.  Pellizzoni, Mereghetti \& De Luca \cite{pelli02} image a H$\alpha$ nebula possibly associated with the central X-ray source in SNR G266.2--1.2.  Gaensler, Jones \& Stappers \cite{gaensler02} image the H$\alpha$ bow shock around PSR J2124--3358.  It is worth noting that the shocks detected in these pulsar wind nebulae comprise both non-radiative and radiative components, based on the absence/presence of lowly-ionized metal lines.

The transverse velocity of the pulsar contributes to its observed period derivative and affects its derived age and magnetic field strength \cite{bell95}.  The separation between the apex of the bow shock and the pulsar is known as the ``stand-off distance'', denoted by $r_w$.  It arises from a balance between the ram pressure of the ISM ($\sim m_{\rm H} n v_{\rm pulsar}^2$) and the pressure from the relativistic pulsar wind ($\sim \dot{E}/4\pi c r^2_w$), where $v_{\rm pulsar}$ is the pulsar velocity and $\dot{E}$ is its spin-down luminosity.  By measuring $v_{\rm pulsar}$ and $r_w$, the ambient density can be estimated,
\begin{equation}
n \sim \frac{\dot{E}}{4\pi m_{\rm H} c r^2_w v^2_{\rm pulsar}}.
\end{equation}
For example, Jones, Stappers \& Gaensler estimate $n \sim 0.25$ cm$^{-3}$ for B0740--28.

Despite establishing the presence of BDS emission in these optical bow shocks, the broad H$\alpha$ line has never been robustly detected in any pulsar wind nebula.

\subsection{High-Redshift Galaxies}

Heng \& Sunyaev \cite{heng08} reason that the same $\sim 1000$--10000 km s$^{-1}$ BDSs --- produced by SNRs in the Sedov-Taylor phase --- that produce broad Balmer lines should also generate luminous (by factors $\sim 10$--100) Ly$\alpha$ lines in young, high-redshift galaxies.  The broad wings\footnote{Heng \& Sunyaev termed these contributions to be ``non-thermal'', but this is not meant to be associated with particles that produce non-thermal X-ray and radio emission.  Rather, it is to indicate hydrogen line emission that is not narrow and typically associated with recombination.} of the Ly$\alpha$ line allow it to escape the local optical depth of the SNR, as demonstrated by the Ly$\beta$ detections of BDSs by Ghavamian et al. \cite{ghava07b}.  At redshifts of $z \sim 2$, the ultraviolet Ly$\alpha$ line is red-shifted into the optical, thus allowing it to avoid Galactic extinction.  If multiple Balmer and Lyman lines are detected from these distant galaxies, they can be used to estimate the average shock velocities and $\beta$-values of the constituent SNRs.

Shapiro et al. \cite{shapiro09} use high-resolution integral field unit spectrographs on 8--10 m-class telescopes to obtain near-infrared spectra of 47 galaxies at redshifts of $z \sim 2$.  Four of these galaxies have previously-identified active galactic nuclei (AGN).  Stacking the spectra of the (putative) non-AGN galaxies reveals a broad component to the H$\alpha$ line with $W \gtrsim 1500$ km s$^{-1}$; in comparison, the 4 galaxies with AGN have a broader ($\sim 3000$ km s$^{-1}$) and stronger (by a factor $\sim 3$) H$\alpha$ line.  In stacked spectra from 7 well-resolved galaxies, the strength of the broad H$\alpha$ line is observed to be greater near the galactic centers ($<3$ kpc) than in the outer regions (3--15 kpc).

By grouping the 47 galaxies into three stellar mass bins ($M_\star < 2 \times 10^{10} M_\odot$, $M_\star = 2$--$7 \times 10^{10} M_\odot$ and $M_\star > 7 \times 10^{10} M_\odot$), Shapiro et al. demonstrate that the strength and width of the broad H$\alpha$ line is correlated with $M_\star$, which can be interpreted in two ways.  Higher stellar masses stem from higher star formation rates --- it is plausible that galactic winds driven by $\sim 10^4$ SNRs in each galaxy are powering the broad H$\alpha$ emission via the creation of broad neutrals in BDSs, as suggested by Heng \& Sunyaev.  Such a mechanism is energetically consistent with the measured luminosities ($\sim 10^{41}$--$10^{42}$ erg s$^{-1}$) because the ISM in these $z \sim 2$ galaxies is $\sim 10$--30 times denser, thereby boosting the broad H$\alpha$ luminosity by factors $\sim 100$--1000.

The second possibility is that the broad H$\alpha$ emission originates from the broad line regions around unobscured AGN.  Higher stellar masses imply higher bulge and (possibly) black hole masses.  By using an empirical mass-luminosity relation to estimate the black hole masses (to within a factor $\sim 3$), Shapiro et al. show that the black holes are about an order of magnitude \emph{less} massive (relative to their stellar bulge masses) than local elliptical and bulge-dominated spiral galaxies.  This is interpreted as evidence for rapid bulge formation in $z \sim 2$ galaxies preceding a few Gyr of super-massive black hole assembly.

The data obtained by Shapiro et al. do not allow one to robustly distinguish between these two possibilities.

\subsection{Symbiotic Stellar Systems}

Symbiotic stars are binary systems typically comprising a white dwarf (WD) and a cool giant star.  Both stars lose mass via winds and the colliding winds lead to complex spectral behavior at all wavelengths from the radio to X-ray.  CH Cygni is perhaps the most studied of symbiotic stars \cite{contini09}.  In 1977, it produced a powerful outburst that lasted until 1986 --- bipolar radio and optical jets appeared towards the end.  A mysterious Ly$\alpha$ line with width $\gtrsim 4000$ km s$^{-1}$ appeared in 1985; during the quiescent phases that followed (1990--1991, 1995), it disappeared along with the hot ultraviolet and optical continuum, then re-appeared during the 1992--1995 active phase (albeit with its line strength weakened by a factor $\sim 2$--3).  Ultraviolet spectra from the {\it International Ultraviolet Explorer} show that the Ly$\alpha$ line is by far the most prominent, outshining the singly- to triply-ionized metal lines by at least an order of magnitude.

Contini, Angeloni \& Rafanelli \cite{contini09} suggest that broad Ly$\alpha$ emission is generated in a BDS from the WD that propagates \emph{away} from CH Cygni (as opposed to the WD wind that collides with the wind from the giant star).  The contemporaneous appearance of the [O~{\sc iii}] $\lambda 5007$ line limits the ambient density to $n\lesssim 10^4$ cm$^{-3}$ in order to be consistent with avoiding collisional de-excitation.  If the [O~{\sc iii}] and Ly$\alpha$ emission have the same origin, then they cannot be from the region in between the WD and giant star, which has densities $\sim 10^8$--$10^9$ cm$^{-3}$.

Selvelli \& Hack \cite{sh85} measure the Ly$\alpha$ flux to be about $2.16 \times 10^{-10}$ erg cm$^{-2}$ s$^{-1}$.  Since the distance to CH Cygni is 270 pc, the Ly$\alpha$ luminosity is about $2 \times 10^{33}$ erg s$^{-1}$.  The theoretically-expected value is
\begin{equation}
L_{\rm bLy\alpha} \approx 4 \pi R^2_{\rm BW} n\left(1 - f_{\rm ion} \right) v_s \epsilon_{\rm bLy\alpha} E_{\rm Ly\alpha},
\end{equation}
where $R_{\rm BW} \approx 2.25 \times 10^{17}$ cm is the blast wave radius and $n(1-f_{\rm ion}) \approx 0.2$ cm$^{-3}$ is the pre-shock neutral density \cite{contini09}.  The energy of a Ly$\alpha$ photon is $E_{\rm Ly\alpha} \approx 10.2$ eV, and we take the number of broad Ly$\alpha$ photons produced per ionization to be $\epsilon_{\rm bLy\alpha} \approx 1$.  If we take the reported lower limit for the line width to be the shock velocity (without correcting for bulk velocity broadening), then we get $L_{\rm bLy\alpha} \approx 8 \times 10^{32}$ erg s$^{-1}$.  Therefore, the broad Ly$\alpha$ line detected in CH Cygni may be energetically consistent with being from a BDS, although further investigation is required.

\subsection{Supernovae}

Komossa et al. \cite{komossa09} detect H$\alpha$ and H$\beta$ lines with complex structures from a transient event in the non-active galaxy SDSSJ095209.56+214313.3.  These line complexes are respectively fitted by 7 and 5 Gaussians: 1--3 components for the narrow H$\alpha$ and [N~{\sc ii}] $\lambda 6548, 6584$ lines; 2 narrow ``horns''; and 2 broad H$\alpha$ line components.  For both H$\alpha$ and H$\beta$, they report $W \approx 2100$ km s$^{-1}$; the line intensities relative to [O~{\sc iii}] $\lambda 5007$ are 9.7 and 0.79, respectively.  

Using the ratio of the [S~{\sc ii}] $\lambda 6716,6731$ doublet, the ambient density is estimated to be $n \sim 10^2$ cm$^{-3}$.  However, the [O~{\sc iii}] $\lambda 4363,5007$ doublet yields $n \sim 10^7$ cm$^{-3}$, clearly indicating that the oxygen lines are originating from a different region.  Adopting $n \gtrsim 10^2$ cm$^{-3}$ implies that the size of the region associated with narrow H$\beta$ emission is $\lesssim 50$ pc.

It is uncertain if this transient event is indeed a SN.  However, one can describe some characteristics of this event if interpreted in the context of a BDS associated with a SN explosion.  Mysteriously, the Balmer decrement for the broad lines is $\gtrsim 5.7$ and goes up to about 12.2, \emph{higher} than that for the narrow lines (2.6--3.2); the narrow Balmer decrement is consistent with Case B recombination \cite{osterbrock06}.  In the contexts of both dust extinction and Lyman line trapping, it is somewhat difficult to construct scenarios where the broad lines are more susceptible than the narrow ones.

If all of the broad H$\alpha$ emission is due to non-radiative shocks of age $t_{\rm age}$, then using equation (\ref{eq:recombination}) to assert $t_{\rm rec} \gtrsim t_{\rm age}$ implies
\begin{equation}
n \lesssim 5 \times 10^6 \mbox{ cm}^{-3} ~\left(\frac{t_{\rm age}}{10 \mbox{ yrs}}\right)^{-1} \left(\frac{m_i}{m_p}\right)^{0.7} \left(\frac{v_s}{2100 \mbox{ km s}^{-1}}\right)^{1.4},
\end{equation}
if we assume $v_s \sim W$.  If it originates from a BDS, the broad H$\alpha$ is emitted at a distance from the SN of
\begin{equation}
R_{\rm BDS} \gtrsim \sqrt{\frac{L_{\rm bH\alpha} \alpha_{\rm rec} t_{\rm age}}{4 \pi v_s E_{\rm H\alpha} \epsilon_{\rm bH\alpha}}} \sim 2 \mbox{ pc},
\end{equation}
where $L_{\rm bH\alpha} \approx 2 \times 10^{41}$ erg s$^{-1}$ is the mean broad H$\alpha$ luminosity observed \cite{komossa09}.  We take the number of broad H$\alpha$ photons per ionization to be $\epsilon_{\rm bH\alpha} \approx 0.2$.

Indeed, H$\alpha$ lines with widths $\sim 1000$ km s$^{-1}$ are commonly detected in supernovae, but the broadening mechanism is usually assumed to be bulk fluid motion.  A rare ($\lesssim 1\%$ by detected number) and intriguing sub-class are the ``Type Ia/IIn'' supernovae that exhibit spectral features typically associated with thermonuclear explosions alongside prominent hydrogen lines.  A striking example is SN 2005gj, which shows both narrow and broad components of H$\alpha$, H$\beta$, H$\gamma$ and He~{\sc i} $\lambda 5876, 7065$ lines \cite{aldering06}.  The broad H$\alpha$ and H$\beta$ lines have widths of $W \sim 2000$ km s$^{-1}$ and luminosities $\sim 10^{40}$ erg s$^{-1}$.

It is tempting to interpret these broad hydrogen lines in the context of BDSs, in part because the lines are observed to be symmetric, somewhat Gaussian and free of distortions typically associated with geometrical effects (e.g., flat-topped or double-horned profiles); the measured Balmer decrements are also large ($\sim 6$).  However, the required mass of the circumstellar medium is prohibitive:
\begin{equation}
M_{\rm CSM} \sim \frac{E_{\rm bH\alpha} m_{\rm H}}{E_{\rm H\alpha} \epsilon_{\rm bH\alpha}} \approx 140 M_\odot \left(\frac{\epsilon_{\rm bH\alpha}}{0.2}\right)^{-1},
\end{equation}
where $E_{\rm bH\alpha} \sim 10^{47}$ erg is the energy of the broad H$\alpha$ line released over $\sim 100$ days, as observed by Aldering et al.  Since Balmer emission is recorded from 11 to 133 days post-explosion, the volume of the circumstellar material is $\sim 10^{50}$ cm$^3$ if the ejecta velocity is taken to be 25000 km s$^{-1}$, implying that the ambient density is $n \sim 10^9$ cm$^{-3}$.  Therefore, even if shocks are responsible for producing the broad H$\alpha$ lines in SN 2005gj, they are probably radiative.

\subsection{T Tauri Stars}

T Tauri stars --- young, luminous and pre-main sequence --- have been observed to emit symmetric H$\alpha$ lines with broad wings often extending to about 500 km s$^{-1}$, somewhat higher than expected estimates for the infall velocity of gas (200--300 km s$^{-1}$).  The H$\alpha$ lines typically have a blue-shifted absorption trough near line center.

Ardila et al. \cite{ardila02} detect H$\alpha$ lines with widths of $W \approx 120$--330 km s$^{-1}$ in DR Tauri, DF Tauri, BP Tauri and RW Auriga (listed in order of ascending $W$).  For these objects, Mg~{\sc ii} $\lambda 2796, 2803$ lines are observed to have spectral profiles that are virtually coincident with H$\alpha$, suggesting a common origin for these lines and ruling out the linear Stark effect as an explanation (which predicts narrower Mg~{\sc ii} than H$\alpha$ lines) .  Curiously, the H$\beta$ lines --- which are only detected in DF Tauri and BP Tauri ---  have widths of about 130--170 km s$^{-1}$, somewhat lower than the H$\alpha$ line widths.  No correlation is found between the accretion rate and the Mg~{\sc ii} line fluxes.  Balmer decrements are not reported by Ardila et al.  The correlation between the H$\alpha$ and Mg~{\sc ii} lines led Ardila et al. to suggest a broadening mechanism that is independent of atomic physics, such as Alfv\'{e}nic turbulence.  

The association between the broad H$\alpha$ lines and BDSs is suggestive but unestablished.  If true, one can study if there exists a correlation between the shock velocity and the abundance of crystalline grains that may have been processed by shock heating.  It is unclear if Mg~{\sc ii} emission (which is from a resonant doublet) can be produced in such a BDS.

\subsection{Extragalactic SNRs}

The optical discovery of individual, extragalactic SNRs with BDSs is unlikely, primarily because non-radiative hydrogen emission is very faint and difficult to detect.  Past searches for distant SNRs were based on the selection criterion of large [S~{\sc ii}]/H$\alpha$ ratios, which violates one of the stated properties of BDSs (see \S\ref{sect:intro}).

\section{The Future}
\label{sect:future}

\subsection{Electron-Ion Temperature Equilibration in Collisionless Shocks}
\label{subsect:equilibration}

\begin{figure}
\begin{center}
\includegraphics[width=\columnwidth]{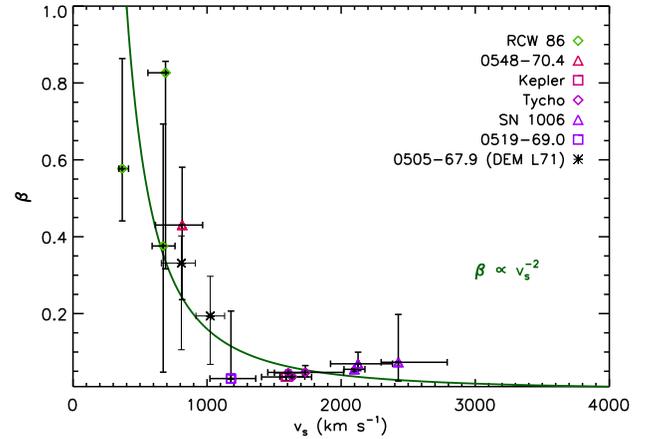}
\end{center}
\vspace{-0.2in}
\caption{Ratio of electron to proton temperatures immediately behind collisionless shock fronts as a function of the shock velocity, obtained via analysis of the two-component H$\alpha$ lines from Balmer-dominated shocks around supernova remnants.  Based on calculations from van Adelsberg et al. \cite{v08}.}
\label{fig:betacurve}
\end{figure}

H$\alpha$ observations of BDSs present a unique opportunity to quantify the degree of electron-proton temperature equilibration in the immediate post-shock regions, potentially yielding insight into the plasma physics of these shock fronts.

Ghavamian, Laming \& Rakowski \cite{ghava07a} use 10 measurements of $W$ and $I_b/I_n$ from four SNRs to derive values of $\beta$ and $v_s$ using the model of Ghavamian et al. \cite{ghava01} (see \S\ref{subsect:models}).  For their fifth SNR, DEM L71, they use the measured $W$ to infer a range of shock velocities and $T_p$ --- $\beta$ is obtained when combined with a {\it Chandra} X-ray estimate of $T_e$.  The curve
\begin{equation}
\beta = 
\begin{cases}
1 & v_s \le 400 \mbox{ km s}^{-1},\\
{\cal B} v^{-2}_s & v_s \ge 400 \mbox{ km s}^{-1},\\
\end{cases}
\end{equation}
appears to fit the data fairly well; they also assume min$\{\beta\} \approx m_e/m_p$.

The Ghavamian, Laming \& Rakowski model (later expanded upon by Rakowski, Laming \& Ghavamian \cite{rakow08}) proposes using lower hybrid waves (see \S\ref{subsect:models_precursors}) to heat the electrons and predicts $\beta \propto v^{-2}_s$, which implies that the electrons are heated to a temperature of
\begin{equation}
T_e = \beta T_p \approx \frac{3 m_p {\cal B}}{16 k} \approx 0.3 \mbox{ keV},
\end{equation}
independent of the shock velocity, suggesting a process occurring within the pre-shock medium rather than the shock front itself.

Detailed modelling by van Adelsberg et al. \cite{v08} shows that
\begin{equation}
\mbox{input: } \left\{W,I_b/I_n\right\} \rightarrow \mbox{output: } \left\{v_s, \beta\right\},
\label{eq:equilibration}
\end{equation}
with a somewhat weak dependence on the pre-shock ion fraction $f_{\rm ion}$ (which they set to 0.5 for their reported results).  Using this approach, they are able to self-consistently derive 14 values of $\beta = \beta(v_s)$ from 7 SNRs (Figure \ref{fig:betacurve}), three more than Ghavamian, Laming \& Rakowski.  Fitting a $\beta \propto v^{-2}_s$ curve to the derived values yields a reduced $\chi^2$ value of about 5, albeit with $\sim 10$--30\% uncertainties in the atomic cross sections (see \S\ref{subsect:atomic}).  Also, they find min$\{\beta\} \approx 0.03 \sim \sqrt{m_e/m_p}$ as opposed to $m_e/m_p$.  Curiously, their derived $\beta$ values hints at a rising trend for $v_s \gtrsim 1000$ km s$^{-1}$, which remains unexplained.

We note that the low values of $\beta \approx 0.06$--0.07, derived for the northwestern filament in SN 1006, are consistent with the findings of Laming et al. \cite{laming96} (see \S\ref{subsect:sn1006}).  Rakowski, Ghavamian \& Hughes \cite{rakow03} combine the analysis of optical Fabry-P\'{e}rot spectra (obtained by Ghavamian et al. \cite{ghava03}) with {\it Chandra} X-ray data to arrive at a similar conclusion for DEM L71 --- their planar shock model yields $\beta \approx 0.01$--0.46 at 4 locations with faster shocks ($v_s \approx 800$--1000 km s$^{-1}$).

Both the studies of Ghavamian, Laming \& Rakowski and van Adelsberg et al. fail to accomodate the low $I_b/I_n$ values measured in DEM L71 (see \S\ref{subsect:large_ibin}) and are unable to extract $\beta$ values from most locations in this remnant.  However, they already rule out the model of Cargill \& Papadopoulos \cite{cp88}, who simulate the heating of the pre-shock medium by shock-reflected ions; their models predict $\beta \sim 0.2$ independent of $v_s$.  It is fair to say that the technique of quantifying electron-proton temperature equilibration from optical studies of BDS is still nascent and requires further refinement, particularly with respect to incorporating percursor physics.

The broad O~{\sc vi} line widths measured by Ghavamian et al. \cite{ghava07b} allow an estimate of the \emph{oxygen ion-proton} temperature equilibration, since
\begin{equation}
\frac{W}{W_{\rm O}} \approx 4 \sqrt{\frac{T_p}{T_{\rm O}}},
\end{equation}
where $W$ is the width of broad Ly$\beta$ in this case; $W_{\rm O}$ and $T_{\rm O}$ are the width of the O~{\sc vi} line and the post-shock temperature of the O~{\sc vi} ion, respectively.  For SNRs 0505--67.9 and 0519--69.0, they measure $W/W_{\rm O} < 4$, implying incomplete ion-proton equilibration.  From three estimated values of $T_{\rm O}/T_p$, they tentatively conclude that there is less ion-proton equilibration at higher shock velocities.  The far-ultraviolet lines alone preclude an estimate of the electron-ion equilibration, because interstellar dust extinction removes the narrow line component.  

Other studies using {\it Far Ultraviolet Spectroscopic Explorer} detections of the $\lambda 1032, 1038$ lines have been performed.  Korreck et al. \cite{korreck04} measure $W_{\rm O} = 2100 \pm 200$ km s$^{-1}$ from SN 1006.  Combined with the Ghavamian et al. \cite{ghava01} value of $W=2290 \pm 80$ km s$^{-1}$ for H$\alpha$, this implies $T_{\rm O}/T_p \approx 13$, less than mass-proportional heating and indicative of poor equilibration.  By contrast, Raymond et al. \cite{ray03} infer that the protons and oxygen ions in northern Cygnus have the same temperatures to within a factor of 2.5.

The microphysics of electron-ion temperature equilibration at collisionless shock fronts is relevant to Sunyaev-Zel'dovich studies of galaxy clusters.  The Compton $y$-parameter, integrated over the area of the cluster, is a proxy for the cluster mass $M_{\rm cluster}$,
\begin{equation}
\int n_e T_e ~dV_{\rm cluster} = {\cal F}\left(M_{\rm cluster}\right),
\label{eq:clusters}
\end{equation}
where $n_e$ is the electron density, $V_{\rm cluster}$ is the volume occupied by the cluster and ${\cal F}={\cal F}(M_{\rm cluster})$ is a power-law function of the cluster mass.  Models that compute $T_e$ typically assume complete equilibration between the electrons and ions.  Rudd \& Nagai \cite{rudd09} show that if partial equilibration is assumed, then the value of $M_{\rm cluster}$ inferred is about 5\% higher.\footnote{This uncertainty is comparable to the $\sim 7$--8\% scatter inherent in using simulations to calibrate the relation in equation (\ref{eq:clusters}).}  More importantly, the uncertainty in the temperature equilibration translates to $\lesssim 30\%$ uncertainties in the electron temperature as a function of distance $r_{\rm cluster}$ from the center of the cluster, introducing an additional source of systematic uncertainty for models attempting to extract the position of the virial shock front from $T_e(r_{\rm cluster})$.

Finally, we note that a review of electron-ion equilibration in collisionless shocks from SNRs can be found in Rakowski \cite{rakow05}.

\subsection{Test Beds for Particle Acceleration in Partially Neutral Media}

The general absence of non-thermal emission at the locations where BDSs are present is consistent with the picture that ion-neutral interactions damp out the Alfv\'{e}n waves necessary for particle acceleration \cite{drury96}, but the dependence of this process on the physical conditions of the shock needs to be further quantified.  In cosmic ray-dominated shocks, it is possible that sufficient electron heating is generated such that no neutral atoms survive to cross the shock, hence rendering ion-neutral damping ineffective/irrelevant.  BDSs represent a unique opportunity to test this hypothesis in more detail.

More interestingly, the locations in the remnants of SN 1006 and Tycho where non-thermal and BDS emission are correlated serve as valuable test beds for models of turbulently-amplified magnetic fields in partially neutral media \cite{bt05,reville07}, and therefore deserve further scrutiny.  

\subsection{Measuring Injection Energy from Face-On Balmer-Dominated Shocks}
\label{subsect:cr_injection}

Raymond et al. \cite{ray83} and Raymond \cite{ray91} suggest a search for broad H$\alpha$ emission from face-on BDSs as a way of testing models for shock acceleration of cosmic rays.  By comparing the velocity shift $\sim 2v_s/(\gamma+1)$ between the narrow and broad hydrogen lines with $W$, one is in essence comparing $v_s$ and $T_i$, respectively.  This provides an estimate of the energy channelled into cosmic ray \emph{ions}.  Specifically, the thermal velocity of the protons is reduced by a factor $1/\sqrt{1+P_{\rm CR}/P_{\rm G}}$, where $P_{\rm CR}$ and $P_{\rm G}$ are the cosmic ray and gas pressures, respectively \cite{ray08}.  For example, $W$ is expected to be $\sim 30\%$ lower if half of the energy goes into non-thermal particles.  Quantifying the presence of cosmic ray ions allows us to test the Rakowski, Ghavamian \& Laming \cite{rakow08} precursor model of lower hybrid wave heating (see \S\ref{subsect:models_precursors}).

Such face-on shocks are extremely difficult to observe, because their brightness is not enhanced by limb brightening like in the case of edge-on shocks.  To date, no reliable catalogues of face-on BDSs exist.  However, recent work by Helder et al. \cite{helder09} combines X-ray proper motion measurements with a detection of the broad H$\alpha$ line from an edge-on BDS in RCW 86 to infer that a substantial fraction of the shock energy is being channelled into cosmic rays.  By empirically constraining the parameter space of $\epsilon_{\rm CR}$ versus $w_{\rm CR}$ --- respectively the fraction of escaping energy associated with cosmic rays (relative to the kinetic energy of the shock) and the ratio of non-thermal to the total post-shock pressure --- they infer $w_{\rm CR} \gtrsim 50\%$.

\subsection{Balmer-Dominated Shock Models with Cosmic Ray Precursors}

There is a lack of models for cosmic ray-mediated shocks in partially neutral gas where $l_{\rm gyro} < l_{\rm CR} < l_{\rm atomic}$ and $v_s \sim 1000$ km s$^{-1}$.  The observations of unusually small and large values of $I_b/I_n$ and $W_0$, respectively, in several SNRs --- as well as the correlation between optical and non-thermal X-ray emission observed in SN 1006 --- provide incentive for constructing these models.

\subsection{H$\alpha$ Line Profiles as Tracers of the Post-Shock Proton Distributions}
\label{subsect:pickup}

Since charge transfer provides a direct mapping between the post-shock protons and broad neutrals, high-precision measurements of the broad H$\alpha$ line can be used to quantify the post-shock proton velocity distribution, particularly if it deviates from being in a Maxwellian distribution.

Raymond, Isenberg \& Laming \cite{ray08} study the effects of ``pick-up ions'' --- scrutinized extensively in the solar wind --- which in the context of BDSs are protons created when neutral atoms pass through the collisionless shock front, become ionized downstream and suddenly ``see'' the ambient magnetic field.  The protons gyrate around the field lines if the associated time scale, $t_{\rm gyro} \sim 2\pi m_p c/e B_{\rm ISM}$, is less than the proton-proton temperature equilibration time $t_{pp}$,
\begin{equation}
B_{\rm ISM} > 0.33 ~\mu\mbox{G} ~\left(\frac{v_s}{1000 \mbox{ km s}^{-1}} \right)^{-3} \left(\frac{n}{1 \mbox{ cm}^{-3}} \frac{{\cal C}}{10} \right),
\end{equation}
where we note that $t_{pp} = \sqrt{m_e/m_p} t_{ep}$ with $t_{ep}$ being the time scale for electron-proton temperature equilibration (equation [\ref{eq:tei}]).

The gyration of these protons forms an unstable ring in three-dimensional velocity space, which then evolves towards isotropy via scattering and self-interaction.  In their study, Raymond, Isenberg \& Laming assume that the end result is a bi-spherical distribution \cite{wz94}.  Pick-up ions introduce non-Maxwellian components to the proton distribution function --- specifically, the contributions occur near the broad H$\alpha$ line center.  Assuming that the broad H$\alpha$ line exactly traces the proton distribution --- an assumption that breaks down for $v_s \gtrsim 2000$ km s$^{-1}$ shocks because charge transfer is no longer the dominant atomic process (see \S\ref{subsect:atomic}) --- Raymond, Isenberg \& Laming find that the shock velocity can be under-estimated by as much as 15\%.  

If the magnetic field is laminar, then the broad H$\alpha$ line is generally asymmetric about line center and is offset by a velocity shift\footnote{This has the implication that edge-on shocks may have a finite velocity shift between the centroids of the narrow and broad H$\alpha$ lines, and is degenerate with the interpretation that the shift yields the angle between the line of sight and the vector normal to the plane of the shock.} from the narrow H$\alpha$ line center.  If $\theta_{\rm B}$ is the angle between the flow across a plane-parallel shock and the magnetic field lines, then the shift varies from zero at $\theta_{\rm B}=0$ to $\sim v_{\rm A}$ at $\theta_{\rm B}=90^\circ$, where $v_{\rm A} \sim 10^{-3}$--$10^{-2} v_s$ (equation [\ref{eq:alfven_mach}]) is the Alfv\'{e}n velocity.  If the magnetic field is turbulent, then the broad H$\alpha$ line is a smeared-out composite of asymmetric lines with a range of $\theta_{\rm B}$ values --- one can in principle extract information about the nature of the turbulence from careful analysis of the line.  Turbulent diffusion dominates thermalization when the associated length scale is
\begin{equation}
l_{\rm turb}  < 2 \times 10^{11} \mbox{ cm} ~\left(\frac{v_s}{1000 \mbox{ km s}}\right)^4 \left(\frac{n}{1 \mbox{ cm}^{-3}} \frac{{\cal C}}{10} \right)^{-1}.
\end{equation}
Generally, both scenarios imply that \emph{information on the magnetic field geometry can be obtained from high-precision studies of the broad H$\alpha$ line.}

The Raymond, Isenberg \& Laming study relaxes the common assumption that newly-created protons are instantly assimilated into the thermal population.  This assumption is especially suspect when the pre-shock neutral fraction is high and the energy of the thermal, post-shock protons is non-negligibly reduced by the act of assimilation \cite{lr96}.  For example, the predicted non-Gaussian contributions to the broad H$\alpha$ line will be much easier to see in Tycho's SNR ($f_{\rm ion} \approx 0.15$) rather than in the remnant of SN 1006 ($f_{\rm ion} \approx 0.9$).  Since the broad H$\alpha$ line is emitted from a region of varying pick-up ion fraction, the creation of pick-up ions should be included in future models of BDSs.

If cosmic ray \emph{protons} are present in the distribution, it is conceivable that they will introduce a third, very broad component to the H$\alpha$ line.  While expected to be exceedingly difficult to detect, its quantification will directly yield the fraction of non-thermal protons by energy.

\subsection{Detecting Polarization in H$\alpha$ Lines}

Laming \cite{laming90} suggests a search for linear polarization in the narrow H$\alpha$ line, which arises when the bulk velocity of the post-shock flow is comparable to the thermal velocities of the electrons or protons.  The polarization arising from the protons is always expected to be present, while the contribution from the electrons decreases as they heat up.  

Using the $f_{\rm eq}$ parametrization (see equation [\ref{eq:feq}]) as introduced by Ghavamian et al. \cite{ghava02} (see their equations [1] and [2]), the electron and proton temperatures in a pure hydrogen gas are:
\begin{equation}
\begin{split}
&T_e \approx \frac{3 m_p v^2_s}{16k} \left(\frac{\beta}{1+\beta}\right),\\
&T_p \approx \frac{3 m_p v^2_s}{16k} \left(\frac{1}{1+\beta}\right).\\
\end{split}
\end{equation}
The respective thermal velocities (equation [\ref{eq:vthermal}]) are then:
\begin{equation}
\begin{split}
&v_{{\rm th},e} \approx \frac{3 v_s}{4} \sqrt{\frac{m_p}{m_e}\left(\frac{\beta}{1+\beta}\right)},\\
&v_{{\rm th},p} \approx \frac{3 v_s}{4} \frac{1}{\sqrt{1+\beta}}.\\
\end{split}
\label{eq:vthermal2}
\end{equation}
Since the bulk velocity of the post-shock fluid is $\Delta v \sim 3 v_s/4$ (with respect to the pre-shock gas), the thermal velocities of the electrons and protons are both comparable to $\Delta v$ when $\beta \ll 1$.  In the case of $\beta = 1$, the protons still have $v_{{\rm th},p} \sim \Delta v$, but the electrons now have $v_{{\rm th},e} \sim 30 \Delta v$.

The pre-shock hydrogen atoms flowing across the collisionless shock front ``see'' an anisotropic distribution of electrons or protons when the respective thermal velocity is $\sim \Delta v$.  If $v_{{\rm th},e} \gg \Delta v$, then the electronic contribution to the polarization is washed out by the random motions of the electrons.  In Case A conditions, Laming estimates that polarizations of up to 10\% may be present for BDSs with $v_s \approx 2000$ km s$^{-1}$ and $\beta = 0$ or 1; the polarization in Case B conditions, for realistic $\beta$ values, is expected to be under 1\%.  The broad H$\alpha$ line is expected to be much less polarized, because the broad neutrals are moving at nearly the same bulk velocity as the post-shock electrons/protons, and thus ``see'' nearly isotropic electron/proton distributions.

If the precursor responsible for broadening the narrow lines (see \S\ref{subsect:broad_narrow_line}) does not induce polarization, then the two narrow line contributions can be distinguished.  More accurate calculations of the polarization present in the narrow H$\alpha$ lines are contingent upon obtaining better cross sections for proton-hydrogen atom excitation (see \S\ref{subsect:scarcity}).  No successful search for polarization in the narrow H$\alpha$ lines from BDSs has ever been reported.

\subsection{The Case for Integral Field Unit Spectroscopy}
\label{subsect:ifu}

Until now,\footnote{We note that Ghavamian et al. \cite{ghava03} observed DEM L71 with a Fabry-P\'{e}rot imaging spectrograph, albeit at sparsely-sampled resolutions.} spectra of BDSs have been obtained using slit spectroscopy or ``two-dimensional'' spectra --- each vertical position along the slit has its own associated spectrum.  Integral field unit (IFU) spectroscopy yields a ``data cube'': one spectral and two spatial dimensions.  In essence, one obtains images of an object at hundreds of different wavelengths (e.g., SNR 1987A by Kj{\ae}r et al. \cite{kjaer07}).  Its strongest advantage over slit spectroscopy is the increased amount of spatial information.  For example, IFU spectroscopy will be useful for studying the spatial variation of the widths and intensities of the narrow and broad H$\alpha$ lines in a resolved shock filament.

\vspace{0.25in}

\textit{This review was inspired by, and dedicated to, the 60th birthday celebration of Roger Chevalier at Caltech.  I acknowledge generous support by the Institute for Advanced Study, the Frank \& Peggy Taplin Membership, NASA grant NNX08AH83G and NSF grant AST-0807444.  I wish to thank Dick McCray, Roger Blandford, Glenn van de Ven, Rashid Sunyaev, Jack Hughes, Steve Reynolds, Matt van Adelsberg, Carles Badenes, Jacco Vink, Doug Rudd, Parviz Ghavamian, Mike Shull, Bruce Draine and Anatoly Spitkovsky for stimulating conversations concerning shocks as well as atomic and plasma physics.}  

\textit{I am especially thankful to: John Raymond, Cara Rakowski, Chris Hirata, Roger Chevalier and Bryan Gaensler for constructive comments following their careful reading of earlier versions of the manuscript; Gamil Cassam-Chena\"{i}, Eveline Helder and Bryan Gaensler for kindly providing sample images and spectra; the referee, Martin Laming, for a thoughtful report and useful conversations.}

\textit{I am grateful to my wife, Stefanie Hetzenecker, for her unwavering love, patience and support.}

\end{document}